\documentclass[10pt,twocolumn,superscriptaddress,aps,pra,tightenlines]{revtex4-2}

\usepackage[utf8]{inputenc}
\usepackage{amsmath,amsfonts,amssymb,amsthm,fullpage,color,graphicx,xcolor,physics,bm,tabularx,booktabs}
\newcommand{\multiline}[1]{%
  \begin{tabularx}{\dimexpr\linewidth-\ALG@thistlm}[t]{@{}X@{}}
    #1
  \end{tabularx}
}

\usepackage{tikz}
\usetikzlibrary{quantikz}

\usepackage[colorlinks]{hyperref}

\newcommand{\eab}{\textsf{EEL}}
\newcommand{\emeab}{\textsf{EMEEL}}
\newcommand{\sysb}{\textsf{EFL}}
\newcommand{\auxb}{\textsf{AuxEFL}}

\newcommand{\coleq}{\mathrel{\mathop:}\nobreak\mkern-1.2mu=}
\newcommand{\eqcol}{\mkern-1.2mu=\mathrel{\mathop:}\nobreak}

\newcommand{\mc}{\mathcal}

\newcommand{\mr}{\mathrm}
\newcommand{\mbb}{\mathbb}

\newcommand{\Pn}{{\sf P}^n}

\newcommand{\pt}{\mathrm{pt}}

\usepackage{algorithm}  
\usepackage[noend]{algpseudocode}  

\begin{document}
\title{Entanglement-enhanced learning of quantum processes at scale}
\author{Alireza~Seif}
\thanks{These authors contributed equally to this work.}
\affiliation{IBM Quantum, IBM T.J. Watson Research Center, Yorktown Heights, NY, USA}
\author{Senrui~Chen}
\thanks{These authors contributed equally to this work.}
\affiliation{Pritzker School of Molecular Engineering, University of Chicago, Chicago, IL, USA}
\author{Swarnadeep Majumder}
\affiliation{IBM Quantum, IBM T.J. Watson Research Center, Yorktown Heights, NY, USA}
\author{Haoran~Liao}
\affiliation{%
Department of Physics, University of California, Berkeley, CA 94720, USA
}%
\author{Derek~S.~Wang}
\affiliation{IBM Quantum, IBM T.J. Watson Research Center, Yorktown Heights, NY, USA}
\author{Moein~Malekakhlagh}
\affiliation{IBM Quantum, IBM T.J. Watson Research Center, Yorktown Heights, NY, USA}
\author{Ali~Javadi-Abhari}
\affiliation{IBM Quantum, IBM T.J. Watson Research Center, Yorktown Heights, NY, USA}
\author{Liang~Jiang}
\affiliation{Pritzker School of Molecular Engineering, University of Chicago, Chicago, IL, USA}
\author{Zlatko~K.~Minev}
\affiliation{IBM Quantum, IBM T.J. Watson Research Center, Yorktown Heights, NY, USA}
\begin{abstract} 
Learning unknown processes affecting a quantum system reveals underlying physical mechanisms and enables suppression, mitigation, and correction of unwanted effects. Describing a general quantum process requires an exponentially large number of parameters. Measuring these parameters, when they are encoded in incompatible observables, is constrained by the uncertainty principle and requires exponentially many measurements. However, for Pauli channels, having access to an ideal quantum memory and entangling operations allows encoding parameters in commuting observables, thereby exponentially reducing measurement complexity. In practice, though, quantum memory and entangling operations are always noisy and introduce errors, making the advantage of using noisy quantum memory unclear. To address these challenges we introduce error-mitigated entanglement-enhanced learning and show, both theoretically and experimentally, that even with noise, there is a separation in efficiency between learning Pauli channels with and without entanglement with noisy quantum memory. We demonstrate our protocol's efficacy in examples including hypothesis testing with up to 64 qubits and learning inherent noise processes in a layer of parallel gates using up to 16 qubits on a superconducting quantum processor. Our protocol provides accurate and practical information about the process, with an overhead factor of $1.33 \pm 0.05$ per qubit, much smaller than the fundamental lower bound of 2 without entanglement with quantum memory. Our study demonstrates that entanglement with auxiliary noisy quantum memory combined with error mitigation considerably enhances the learning of quantum processes. 
\end{abstract}
\maketitle

\section{Introduction}
Understanding and characterizing unknown quantum processes are crucial for advancing physics and developing engineered quantum systems. A detailed knowledge of noise processes enables quantum computers to perform useful tasks with the help of quantum error mitigation~\cite{temme2017error,van2023probabilistic,kim2023evidence} in the near-term and through quantum error correction in the future~\cite{lidar2013quantum}. Describing a general quantum process becomes increasingly complex as the system size grows, due to the exponential growth of Hilbert space. To learn the parameters describing a process, one measures the physical observables that encode them. However, these observables often do not commute, and due to the constraints imposed by the uncertainty principle, learning them requires an exponentially large number of measurements. 

Counterintuitively, entanglement, which is one of the main factors in increasing complexity, can enhance our ability to learn properties of physical systems if used cleverly. For example, entangling quantum sensors provides a quadratic advantage in their sensitivity~\cite{RevModPhys.89.035002,PhysRevLett.96.010401}. Entangled measurements in quantum tomography can reduce its cost to a degree~\cite{PhysRevLett.90.193601,mohseni2008quantum,PhysRevLett.129.133601,PhysRevLett.86.4195}. Recently, it has been shown that entanglement with quantum memory can in  theory provide exponential advantage in learning quantum processes~\cite{aharonov2022quantum,chen2022quantum,chen2022exponential,huang2022quantum,caro2022learning}. However, in practice, one might suspect that adding even more fragile entanglement and unavoidably-faulty auxiliary components to an already noisy system should limit the entanglement-based advantage~\cite{PhysRevLett.79.3865,escher2011general,demkowicz2012elusive}. In some classification and hypothesis testing tasks, this advantage can partially survive in experiments~\cite{huang2022quantum}.  But beyond simple proofs of principle, can this advantage be observed in practical real-world experiments?

\begin{figure*}[t]
    \centering    \includegraphics[width=\textwidth]{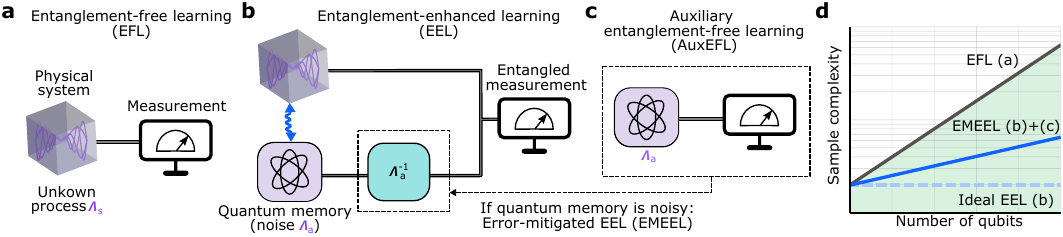}
    \caption{\textbf{Different scenarios for learning an unknown quantum process.} \textbf{a,} Entanglement-free learning (\sysb{}) with the direct measurement of the process $\Lambda_s$ (a Pauli channel in this work).  \textbf{b,} Entanglement-enhanced learning (\eab{}) using a quantum memory. This strategy, through entanglement with an ideal noiseless quantum memory, provides an exponential advantage in the number of samples required for learning $\Lambda_s$. In practice, quantum memory can be noisy (affected by another unknown process $\Lambda_a$). \textbf{c,} Noise on quantum memory, $\Lambda_a$, can be learned efficiently using direct entanglement-free learning on auxiliary qubits (\auxb{}). This noise can be simplified to a generalized depolarization channel, making \auxb{} considerably simpler and more efficient than \sysb{}. We can then apply error mitigation in panel (b) and apply $\Lambda^{-1}_a$ to recover correct results with possibly an exponential overhead using error-mitigated \eab{} (\emeab{}). \textbf{d,} The sample complexity of \sysb{} scales exponentially, whereas  that of \eab{} remains independent of system size.  Although \emeab{} also has exponential sample complexity, it can break \sysb{}'s fundamental lower bound and achieve better scaling by leveraging entanglement with quantum memory and error mitigation, provided the noise on the quantum memory is weak.}
    \label{fig:schematic}
\end{figure*}
Here, we demonstrate that the advantage of using entanglement with quantum memory for learning physical processes can persist even in the presence of noise. This is achieved by designing a protocol that efficiently mitigates errors in the noisy auxiliary quantum memory (see Fig.~\ref{fig:schematic}). Specifically, we focus on the practical case of learning Pauli channels without any assumptions about their structure. Pauli channels are now routinely used in state-of-the-art benchmarking~\cite{harper2020efficient,erhard2019characterizing}, error mitigation~\cite{temme2017error,van2023probabilistic,kim2023evidence,ferracin2022efficiently}, and error correction~\cite{aliferis2008fault, chubb2021statistical} applications. Moreover, in the ideal scenario, there are rigorous guarantees of exponential separation between learning with and without quantum memory for Pauli channels~\cite{chen2022quantum,chen2024tight,chen2023futility,fawzi2023lower}. We show, both theoretically and experimentally, that despite the generally exponential overhead from error mitigation, our protocol not only rigorously improves the sample complexity of learning but also breaks the fundamental limit on the number of learning samples needed without access to entanglement with a quantum memory.

\section{Learning Pauli channels with and without quantum memory}
Pauli channels are crucial in quantum information processing because they provide a practically relevant model for errors in quantum computers~\cite{nielsen2010quantum,flammia2020efficient}. Physically, Pauli channels describe a probabilistic mixture of Pauli operations. Their practical relevance is further motivated by the fact that arbitrary quantum channels can be transformed into Pauli channels through Pauli twirling, which involves averaging over the application of random Pauli gates before and after the channel~\cite{wallman2016noise,hashim2020randomized,ware2021experimental}. 

A Pauli channel $\mathcal{E}$ on $n$ qubits can be fully specified by its $2^n$ Pauli fidelities $\lambda_k$ corresponding to Pauli operator $P_k$, i.e.,  $\mathcal{E}( \rho)=\sum_k \lambda_k P_k \Tr(\rho {P}_k)/4^n$. These fidelities can be learned by preparing and measuring a quantum system in different bases~\cite{flammia2020efficient}. However, it has been shown recently that such a scheme requires $\Omega(2^n/\varepsilon^2)$ measurements to estimate every $\lambda_k$ to $\pm\varepsilon$ precision with high probability~\cite{chen2024tight,chen2023futility}. This lower bound holds even if one is allowed to concatenate arbitrarily many copies of the Pauli channel and  control operations before doing one measurement. We refer to this general class of protocols that only utilize the qubits directly affected by the channel as entanglement-free learning (\sysb{}), see Fig.~\ref{fig:schematic}a. Note that with this terminology we still allow for entangling operations acting on the system alone, but exclude entanglement between system and auxiliary memory qubits.

Access to quantum memory, however, significantly alters the resource requirements. Specifically, we entangle each system qubit with an auxiliary qubit in the Bell basis, i.e., $\ket{\Phi_1}=1/\sqrt{2}(\ket{00}+\ket{11})$. The state of  $n$ such pairs is then given by $\ket{\Phi_n}={2^{-n/2}}\sum_{i=0}^{2^{n-1}} \ket{ii}$, where the first and second index correspond to the $2^n$ computational basis states of $n$ system and auxiliary qubits, respectively. Applying the Pauli channel $\mathcal{E}$ to the system and  the identity channel $\mc I$ to the auxiliary qubits prepares
\begin{equation}\label{eq:choi}
    (\mathcal{E}\otimes \mathcal{I})(\ketbra{\Phi_n}) = \sum_k \lambda_k P_k \otimes P_k^{\sf T}/4^n\;, 
\end{equation}
which is also known as the Choi state of the channel~\cite{choi1975completely}. We then observe that we can measure all $P_k \otimes P_k^{\sf T}$ simultaneously and obtain the corresponding $\lambda_k$ as they commute, i.e., $[P_k \otimes P_k^{\sf T}, P_j \otimes P_j^{\sf T}]=0$ for all $P_k,P_j$. 
This can be done via the Bell basis measurement, akin to a multi-qubit version of superdense coding~\cite{bennett1992communication}. We refer to this scheme as entanglement-enhanced learning (\eab{}), see Fig.~\ref{fig:schematic}b.
Its sample complexity (see SM Sec. II) is $O(1/\varepsilon^2)$~\cite{chen2022quantum}. This scheme theoretically offers an exponential advantage over \sysb{}, making it a promising strategy for learning Pauli channels. However, achieving this advantage in practice is challenging. Errors in state preparation and measurement (SPAM) and entangling operations can increase sampling overhead and diminish the advantage. Additionally, the scheme depends on having access to perfect auxiliary qubits; errors on these qubits can bias the results and lead to incorrect parameter estimates.

 \begin{figure*}[t]
    \centering
    \includegraphics[width=\textwidth]{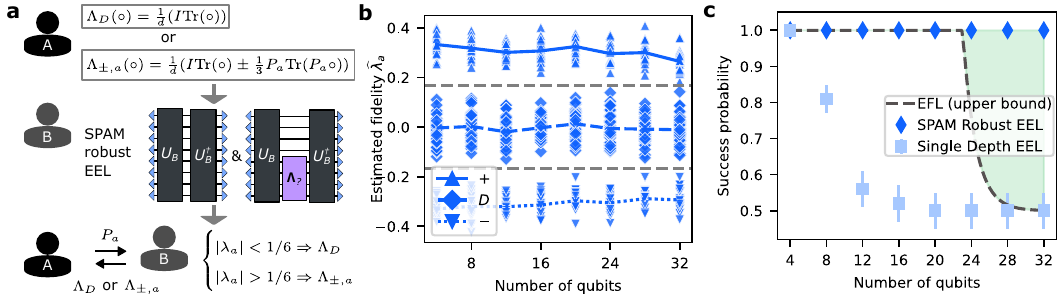}
    \caption{
    \textbf{Hypothesis testing. }%
    \textbf{a,} Schematics of the task and the SPAM robust entanglement-enhanced learning (\eab{}) protocol. The unitary operations $U_B$ and $U_B^\dagger$ correspond to state-preparation and measurement in the Bell basis between system and auxiliary qubit pairs, respectively. 
    See Sec.~\ref{sec:distinguish} for details. 
    \textbf{b,} Estimates for $\lambda_a$ from SPAM robust \eab{}.
    The diamond, triangle, and inverted triangle data points correspond to the three types of Pauli channels, $\Lambda_D$, $\Lambda_{-,a}$, and $\Lambda_{+,a}$, in which case the true value of $\lambda_a$ is $0,-1/3,+1/3$, respectively.
    For each system size $n\in[4,8,\cdots,32]$, we sample $a\in{\sf P}^n$ 50 times and plot the estimate for $\lambda_a$ for each type of Pauli channel. The average over $a$ for each type of Pauli channel is shown as solid lines.  
    Clearly, by comparing the estimates with the thresholds $\pm1/6$ (dash line), one can decide the type of underlying Pauli channels with high probability.
    \textbf{c,} Comparison of success probability in distinguishing $\Lambda_D$ with $\Lambda_{\pm,a}$ via single-depth \eab{}, SPAM robust \eab{}, and entanglement-free learning (\sysb{}) upper bound given in Eq.~\eqref{eq:qmf_upper}. The single-depth \eab{} scheme uses $M=5\times10^6$ total measurements, while SPAM-robust \eab{} uses twice that number, $M=5\times10^6$ for SPAM characterization and the same number for Pauli channel learning. While the success probability of single-depth \eab{} quickly converges to $1/2$ as system size increases, SPAM-robust \eab{} retains a success probability of almost $1$ and surpasses the upper bound for \sysb{} for $M=10^7$ measurements, demonstrating a rigorous entanglement enhancement. The error bars and samples are obtained from random sampling of injected Pauli noise on \emph{ibm\_brisbane} (see SM). 
    }
    \label{fig:decision}
\end{figure*}

In the following, we address these challenges by introducing a SPAM-robust protocol and designing an error mitigation technique specifically tailored for such learning tasks, which we refer to as error-mitigated entanglement-enhanced learning (\emeab{}), see Fig.~\ref{fig:schematic}b~and~c. To mitigate errors with \emeab{}, we first perform additional entanglement-free learning experiments on auxiliary qubits (\auxb{}). While both \auxb{} and \sysb{} are entanglement-free strategies, \auxb{} is considerably simpler and more efficient because we can tailor the noise affecting the idle auxiliary qubits. We then use this information to extract the correct error-free fidelities from \eab{} results. Although \emeab{} introduces some additional overhead due to error mitigation, it can surpass the theoretical limit of \sysb{} when noise is weak and achieve a sampling cost that is orders of magnitude smaller. We demonstrate this enhancement through several examples involving distinguishing and characterizing Pauli channels.

\section{Hypothesis testing}\label{sec:distinguish}
\begin{figure*}[t]
    \centering
    \includegraphics[width=\textwidth]{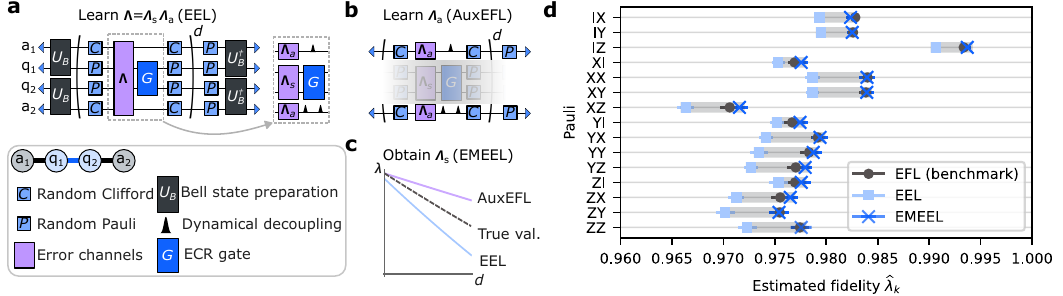}
    \caption{ \textbf{Learning the noise of a two-qubit gate. a,} The quantum circuit for entanglement enhanced learning of noise affecting the quantum gate $G$ applied to qubits $q_1$ and $q_2$, which is chosen to be an echoed cross-resonance (ECR) gate \cite{shedon2016cr, malekakhlagh2020firstprinciples, sundaresan2020reducing, itoko2024three} in the experiment. The gate is applied $d$ times interleaved by twirling layers. Here, $d=0,16,32$. Auxiliary qubits $a_1$ and $a_2$ are prepared in a pairwise maximally entangled Bell state with their neighboring system qubits using $U_B$. The auxiliary qubits are twirled using single qubit Clifford gates (C), while the system qubits are twirled using random Pauli gates (P). In general, the error channel $\Lambda$ affects all the qubits in a correlated way. However, when these errors are coherent, using a simple dynamical decoupling strategy we can decouple the error channel into $\Lambda_a\otimes \Lambda_s$, affecting the auxiliary and system qubits in an uncorrelated way, respectively. \textbf{b,} We learn $\Lambda_a$ by preparing and measuring the auxiliary qubits in a fixed basis while the gate is applied to the system. Because of Clifford twirling, the Pauli fidelities of $\Lambda_a$ only depend on the pattern of the Pauli string, i.e., whether a non-trivial Pauli operator acts on the qubits. \textbf{c, } The Pauli fidelities are extracted from the decay of the expectation value of the output of the circuits in panels a and b as a function of $d$. The \eab{} curve decays faster than the \sysb{} curve. The slope of these curves determine the corresponding Pauli fidelity. When $\Lambda = \Lambda_a \otimes \Lambda_s$, the slope of \eab{} curve is the product of \sysb{} and \auxb{}'s. \textbf{d,} Estimated Pauli fidelities $\hat\lambda_k$   of $\Lambda_s$ for the 15 non-trivial Pauli ($P_k$) operators on two qubits. We measure the fidelities using two independent methods with and without quantum memory. We take \sysb{} as the ground truth. Without error mitigation \eab{} results do not match \sysb{}'s, with their disagreement highlighted in grey.  However, after applying error mitigation we uncover correct fidelities with \emeab{}. Error bars are statistical correspond to 1.96 standard deviation. The experiments were performed on \emph{ibm\_peekskill} with 100 twirls and 1000 shots at each depth.}
    \label{fig:singleeab}
\end{figure*}
We first demonstrate rigorous entanglement enhancement in Pauli channel learning through a hypothesis-testing experiment and highlight the importance of SPAM robustness. In the ideal scenario with noiseless quantum memory, this example has a provable exponential separation in sample complexity between \sysb{} and \eab{} in the ideal case~\cite{PhysRevA.105.032435,chen2024tight},  making it a rigorous benchmarking tool for our experiment. The success probability of \sysb{} decays exponentially with system size. Consequently, if \eab{} experiments, which are inevitably affected by hardware noise, succeed with a higher probability, they provide evidence for entanglement enhancement in this task.

Specifically,  we consider the following game between Alice and Bob of distinguishing random Pauli channels.
Alice first uniformly randomly samples a Pauli $P_k$ and a sign $\pm$. She then chooses one of the following two Pauli channels with equal probability,
\begin{align}
    \Lambda_D(\rho) &= \frac{1}{2^n} \left[I \Tr(\rho)\right]\;, \\ 
    \Lambda_\pm(\rho) &= \frac{1}{2^n} \left[I \Tr(\rho) \pm \frac{1}{3} P_k \Tr(\rho P_k) \right]\;.
\end{align}
Denote her choice as $\Lambda$. Bob is then allowed to probe copies of $\Lambda$ using a certain class of measurement schemes. After Bob's measurements are done, Alice reveals the choice of $P_k$ (but hides the sign $\pm$). Bob is then challenged to guess whether $\Lambda=\Lambda_D$ or $\Lambda=\Lambda_\pm$ by classically post-processing the measurement data he collects.
Ref.~\cite{chen2024tight} shows that without utilizing entanglement with a quantum memory, Bob's average success probability for a given number of measurements $M$ decays exponentially with system size $n$ as (see SM Sec.~II~C)
\begin{equation}\label{eq:qmf_upper}
p_s\le \min(1,\frac12+0.43\times M2^{-n}),\quad\forall n\ge4\;.
\end{equation}
In contrast, using \eab{} with perfect quantum memory and entangled state preparation and measurement, Bob can correctly guess Alice's choice with a constant high probability independent of system size~\cite{chen2022quantum}. Concretely, Bob only needs to compute an estimator $\hat\lambda_k$ such that $|\hat\lambda_k-\lambda_k|<1/6$ with high probability, and decides the channel was $\Lambda_D$ if he obtains $|\hat{\lambda}_k|<1/6$ and $\Lambda_\pm$ otherwise.

In real experiments, however, SPAM errors limit \eab{} and cause its success probability to decay exponentially as they generally push $\hat\lambda_k$ to smaller values.
Fortunately, there are protocols similar to randomized benchmarking~\cite{knill2008randomized,magesan2011scalable,flammia2020efficient,erhard2019characterizing} that remove sensitivity to SPAM errors. These protocols isolate and amplify the fidelity of interest by repeatedly applying the channel. The fidelity can then be extracted from the decay rate of the measured Pauli expectation value as a function of the number of channel applications $d$.  %
This approach is applicable to \eab{} as well. In this example, we use a SPAM-robust protocol with $d=0,1$ to calibrate away SPAM errors and demonstrate a robust entanglement enhancement (see SM Sec. II B).

We experimentally demonstrate the improved success probability of \eab{} compared to \sysb{} by emulating $\Lambda$ through inserting random Pauli gates, using up to 64 qubits (see SM for details of the noise injection protocol). We vary the system size from 4 to 32 qubits and compare the success probability of single-depth \eab{} with SPAM-robust \eab{} and the upper bound for \sysb{} strategies (see Fig.~\ref{fig:decision}). While single depth \eab{} fails to outperform \sysb{}, the SPAM robust protocol succeeds with overwhelming probability. Notably, for $n\geq24$, the success probability of SPAM-robust \eab{} remains constant, exceeding the exponentially small upper bound for protocols without access to quantum memory, demonstrating rigorous entanglement enhancement (green shaded region in Fig.~\ref{fig:decision}c). Note that because of small single-qubit gate errors and their short gate lengths errors on auxiliary qubits do not affect the results for the system sizes considered. The robustness to these small errors is due to the large margin in the decision boundary. However, we expect the success probability on noisy devices to eventually decay as system sizes grow.

\section{Learning noise in quantum operations}

Next, we move beyond emulated noise and learn real noise processes affecting gate operations in quantum processors. We model a noisy multi-qubit Clifford gate as $\widetilde{\mathcal G}=\mathcal G\circ\Lambda$, where $\mathcal G$ is the ideal gate and $\Lambda$ is the noise process, including classical and quantum crosstalk, which is twirled into a Pauli channel~\cite{flammia2020efficient,erhard2019characterizing}. The goal is to utilize entanglement to more efficiently learn the Pauli eigenvalues of $\Lambda$. Since these eigenvalues can only be characterized up to fundamentally unlearnable degrees of freedom~\cite{chen2023learnability}, we focus on learning the symmetrized Pauli eigenvalues~\cite{erhard2019characterizing,hashim2020randomized,van2023probabilistic} (see Methods Sec.~B).

The main challenge in improving sample complexity and obtaining accurate results with entanglement-enhanced schemes is the additional errors introduced by auxiliary qubits during the application of quantum gates on the system. These errors alter the fidelities extracted by \eab{} and bias the results. To address this, we introduce \emeab{}, which incorporates an error mitigation procedure to remove this bias and recover the true fidelities.

The main idea behind \emeab{} is based on the fact that when errors affecting the joint state of the system and auxiliary qubits are uncorrelated, the channel takes a tensor-product form. We use this structure to design a protocol that removes the bias introduced by auxiliary noise, at the cost of an increased variance in our \emeab{} estimates, which we characterize later. In our experiments, we enforce this assumption by applying a context-aware dynamical decoupling sequence on the (idle) auxiliary qubits that removes correlated coherent errors between them and the system~\cite{seif2024suppressing}. %
In these cases, it is possible to find a simple expression for the fidelities extracted by \eab{}. %

Let $\mathcal{E'}$ denote the Pauli noise channel acting on the auxiliary qubits with Pauli fidelities $\lambda'_k$. We find that 
\begin{equation}
(\mathcal{E}\otimes\mathcal{E}')(\ketbra{\Phi}^{\otimes n}) = \sum_k \lambda_k\lambda'_k P_k \otimes P_k^{\sf T}/4^n\;. 
\end{equation}
Therefore, the fidelities extracted by \eab{} are simply scaled by their corresponding auxiliary fidelities. Consequently, we can first learn $\lambda'_k$ without entanglement enhancement and then find the $\lambda_k$ by simply rescaling the noisy \eab{} output. We refer to the former as auxiliary entanglement-free learning (\auxb{}). The combined application of \eab{} (Fig.~\ref{fig:singleeab}a) and \auxb{} (Fig.~\ref{fig:singleeab}b) followed by the post-processing step of rescaling fidelities (Fig.~\ref{fig:singleeab}c) constitute \emeab{}. Note that both the \eab{} fidelities and \auxb{} fidelities can be extracted using the SPAM robust technique outlined earlier. 

While learning $\mathcal{E}'$ using \auxb{} is generally as difficult as the original \sysb{} problem, the noise affecting idle auxiliary qubits can be simplified to a generalized depolarization channel using single-qubit Clifford twirling~\cite{gambetta2012simulrb}. As a result, $\lambda'_k$  depend only on the Pauli pattern, not the specific directions. This allows them to be measured simultaneously, since Pauli operators in the same basis commute (e.g., ${I, Z}^{\otimes n}$).
Note that the same simplification using single-qubit Clifford twirling could not be done to the system due to the existence of multi-qubit gates~\cite{wallman2016noise}. The \emeab{} protocol is explained in details in Methods Sec.~B. More details on the experiments, circuits, and additional data are presented in SM Sec. VI. 

\begin{figure*}[t]
    \centering
    \includegraphics[width=\textwidth]{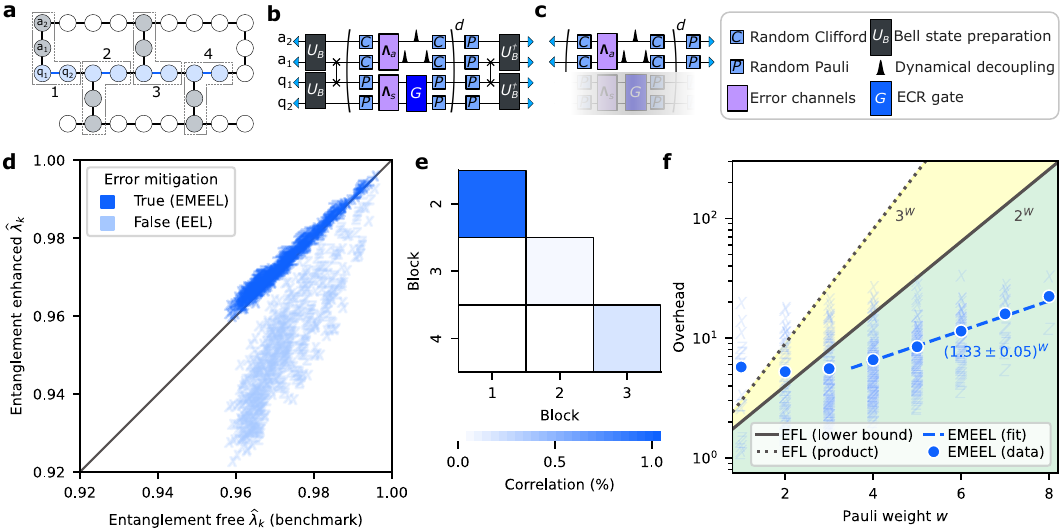}
    \caption{
    \textbf{Learning noise of a layer of parallel gates.} \textbf{a, } The layout of the experiment for learning the noise channel affecting parallel gates in an 8-qubit layer (blue qubits). The layer is divided into 4 blocks of qubits (numbers along the borders). %
    \textbf{b, }
    The quantum circuit for entanglement enhanced learning (\eab{}) of noise  is similar to that of Fig.~\ref{fig:singleeab} for a single block in panel a. In this layout we need a SWAP operation to create a Bell state on $a_{1(2)}$ and $q_{1(2)}$. We use depths $d=0,16,32$, and extract the  fidelity estimates that are robust to state-preparaion and measurement errors from the slope of the decay curves.  %
    \textbf{c,} Noise on auxiliary qubits, $\Lambda_a$, is learned using auxiliary entanglement-free learning protocol (\auxb{}) on those qubits.  %
    \textbf{d, }  The estimated Pauli fidelities $\hat \lambda_k$ with \eab{} using quantum memory and  entanglement-free learning without quantum memory   (\sysb{}) should ideally match (45 degree line) as they correspond to the same physical quantity obtained from two independent methods. However, due to  the bias from auxiliary noise they do not. Error-mitigated entanglement-enhanced learning (\emeab{}) restores the correct values by using information from \auxb{} to mitigate \eab{} estimates. While \emeab{} enables access to all the Pauli fidelities, due to the prohibitive cost of \sysb{} we only benchmark them in the $X$ and $Z$ basis (markers). 
    \textbf{e, } The average correlation [see Eq.~\eqref{eq:corr}] between two-qubit $XX$ and $ZZ$ fidelities of each pairs of blocks. The correlation between adjacent blocks indicates that errors are not local within each block. 
    \textbf{f, } The overhead of \emeab{} compared with a theoretical lower bound for \sysb{} as a function of the Pauli weight $w$, i.e., the number of non-trivial qubits that the operator acts on. The lower overhead of \emeab{}, $(1.33\pm0.05)^w$ from fit  (dashed line) compared to the $2^w$ \sysb{} lower bound (solid black line) and $3^w$ scaling of only using single qubit measurements (dotted black line) demonstrates  entanglement enhancement in sample complexity.
    The experiments were performed on \emph{ibm\_kyiv} with 500 twirls and 500 shots at each depth. 
    }
    \label{fig:multieab}
\end{figure*}
\subsection{Experimental learning of a two-qubit gate noise}
We first demonstrate \emeab{} by characterizing the noise of a two-qubit gate. To validate the results, we additionally perform \sysb{} in various bases and extract all the Pauli fidelities of the noise channel (see Fig.~\ref{fig:singleeab}d). Ideally, the Pauli fidelities from \emeab{} and \sysb{} should match, but \emeab{} achieves this with only 2 measurement settings (one for \eab{} and one for \auxb{}), while \sysb{} requires 9 ($\{X,Y,Z\}^{\otimes2}$). Without error mitigation, the fidelities extracted by \eab{} are smaller than those from \sysb{} with a mean absolute error of ($3.84\pm0.14) \times 10^{-3}$, because noise on auxiliary qubits biases the estimates towards lower values. However, by characterizing this bias with \auxb{}, \emeab{} corrects for it and improves the estimation error by an order of magnitude to $(4.4\pm1.4)\times 10^{-4}$. The success of \emeab{} also validates the tensor product assumption and demonstrates that dynamical decoupling has effectively decoupled the system and auxiliary qubits.

\subsection{Experimental learning of the noise on a layer of two-qubit gates}
While \emeab{} drastically reduces the number of measurement settings from an exponential in $n$ for protocols without memory to 2 (one for $\eab{}$ and another one for $\auxb{}$ as shown in Fig.~\ref{fig:singleeab}b~and~c), the increased variance from additional entangling gates and auxiliary noise increase the number of times we have to measure in each setting. In the next case, we learn the noise  on a parallel layer of 4 two-qubit gates on 8 qubits and answer this question about the overhead and scalability of the protocol (see Fig.~\ref{fig:multieab}). In SM Sec. VI we include additional results using a different layout.

We consider a configuration of parallel gates on a linear chain of qubits (Fig.\ref{fig:multieab}a). This setup on the heavy-hex geometry of our quantum processor requires using SWAP gates to establish entanglement between auxiliary and system qubits (Fig.\ref{fig:multieab}b). These additional gates contribute to SPAM errors, increasing the overhead of \emeab{}, but do not alter the learning protocols (Fig.\ref{fig:multieab}b and c). 

To benchmark the accuracy of \emeab{}, unlike the previous case, we cannot perform \sysb{} for all the Pauli fidelities due to their exponential size. Instead, we measure the fidelities using \sysb{} only in the $X$ and $Z$ bases and compare them with those from \emeab{} (Fig.~\ref{fig:multieab}d). The fidelities from both methods should ideally match. Without error mitigation, \eab{} underestimates the fidelities due to bias introduced by auxiliary noise with a mean absolute error of $(2.081\pm0.004)\times10^{-2}$. With error mitigation, \emeab{} corrects this bias, improving the error by an order of magnitude to $(1.59\pm0.04)\times10^{-3}$. In SM Sec. VI, we show that the small residual bias at lower fidelity values can be improved by using a different layout on the device.

Next, we characterize the correlations in noise between different qubits. Generally, geometric locality can improve the overhead of learning without quantum memory (see e.g. Refs.~\cite{shabani2011esitmation,dasilva2011practical,van2023probabilistic}). For example, if the noise on gates are completely uncorrelated, one could learn the noise affecting each gate independently and efficiently without quantum memory. However, since our entanglement enhanced protocol does not assume locality of errors on the system we can use it to probe the correlations in noise. The presence of correlations limit the applicability of protocols relying on the locality of errors. We characterize the correlation using
\begin{equation}\label{eq:corr}
    \delta_{P} \coleq \left|\lambda_{PPPP} - \lambda_{PPII}\lambda_{IIPP}\right|,\quad P\in\{X,Z\}.
\end{equation}
which quantifies the deviation of errors from a tensor product channel on two pairs of qubits. We then show the average correlation in $X$ and $Z$ bases, i.e., $(\delta_{X}+\delta_{Z})/2$, in Fig.~\ref{fig:multieab}e. 
We observe correlations between adjacent pairs, but do not observe a strong correlation beyond that. We believe that the difference in the strength of correlations between adjacent pairs is related to the direction of applied gate and the dynamical decoupling properties of the gates~\cite{shedon2016cr}.

Finally, we compare the resource cost of \emeab{} and \sysb{}.
Suppose the goal is to estimate all Pauli eigenvalues of Pauli weight up to $w$.
Intuitively, the advantage of $\emeab{}$ is a drastic reduction of state preparation and measurement settings -- only one is needed for $\eab{}$ and another one for $\auxb{}$. 
In contrast, $\sysb{}$ protocols require at least $2^w$ settings using minimal stabilizer covering~\cite{flammia2020efficient} which is experimentally challenging, or at least $3^w$ settings using standard Pauli measurement~\cite{erhard2019characterizing}~\footnote{
Since we are only learning symmetrized Pauli fidelities, one might not need to go through all $3^w$ Pauli basis measurements. However, a lower bound of $\Omega(2^w)$ measurements is still inevitable in this setting~\cite{chen2024tight}.
}.
However, \emeab{} experiences an increased variance in estimating each eigenvalue due to additional SPAM noise and auxiliary QEM. Thus, for \emeab{} to have an enhancement, the reduction in experimental settings must outweigh the variance overhead. 

We first analytically analyze the variance overhead and then compare it with the experimental data. Under reasonable assumptions, this overhead for $\lambda_k$ roughly scales as $(\alpha_k^\sysb/\alpha_k^\eab)^2$,  where $\alpha_k$ is the SPAM Pauli fidelity of the corresponding protocol (see Methods Sec.~C). This overhead scales exponentially with the number of qubits $n$ when SPAM errors are uncorrelated, but should grow much slower than $2^n$ given a reasonable fidelity of the Bell state preparation and measurements. To estimate the overhead, we count the number of two qubit gates, measurements, and idle periods in the protocols. For  \emeab{}, as shown in Fig.~\ref{fig:multieab}b, we have the total of 5 ECR gate in state preparation (2 for Bell state preparation and 3 for SWAP). Additionally, during the SWAP operation two qubits are idle and are affected by T1 and T2 errors. The Bell measurement consists of similar operations with the additional of 4 measurements. In the corresponding \sysb{} protocol, only measurement errors on two system qubits contribute to SPAM errors. Using the reported gate and readout fidelities together with the T1 and T2 time averaged over the qubits involved in the experiment in Table~\ref{tab:fids}), we calculate the average $\alpha_k^\sysb$ and $\alpha_k^\eab$ per qubit and estimate the overhead to be $1.23\pm0.03$. Noting that the major part of the overhead comes from the SWAP errors, we anticipate that utilizing native SWAP gates~\cite{ali2024native} can lead to further improvements. 

\begin{table}[h]\label{tab:fids}
    \centering
    \begin{tabular}{|c|c|}
        \hline
        \textbf{Component} & \textbf{Fidelity (\%)} \\
        \hline
        ECR & $99.03\pm0.11$ \\
        \hline
        T1 &  $99.83\pm0.02$\\
        \hline
        T2 &  $99.39\pm0.17$\\
        \hline
        Readout &  $99.25 \pm 0.17$\\
        \hline
    \end{tabular}
    \caption{The average fidelity of operations corresponding to qubits and gates in Fig.~\ref{fig:multieab}. The T1 and T2 rows correspond to the fidelity of idle operation with those errors for the duration of an ECR gate.}
\end{table}

The overhead extracted from the experimental data corroborates our theoretical prediction. Specifically, our experiments  show that while the overhead grows exponentially, the base of the exponential ($1.33 \pm 0.05$) is much smaller than the theoretical lower bound ($2$) and naive scaling ($3$) for learning without quantum memory (Fig.~\ref{fig:multieab}f), in agreement (within 1.25 standard deviations) with the analytically estimated overhead. These experiments demonstrate that even with noisy quantum memory, entanglement can overcome the theoretical limits on \sysb{} sample complexity, enabling more efficient extraction of useful information about quantum processes in real experiments.

\section{Outlook}
The slower growth of the sampling cost of \emeab{} with system size compared to that of \sysb demonstrates that entanglement with noisy auxiliary quantum memory considerably enhances the learning of quantum processes.  This lower sampling overhead can enable quantum characterization at larger scales without requiring assumptions about locality of the errors on the system. Moreover, the prospect of learning properties of physical systems more efficiently with noisy quantum memory aided by error mitigation opens up possibilities beyond characterizing Pauli channels. For example, identifying symmetries and measuring nonlinear properties of physical systems such as entanglement entropy may also benefit from this technique.  

While we did not make any assumption about the locality of noise in the system qubits, there is an implicit assumption about the locality of interactions between system and auxiliary qubits. This assumption was used in designing dynamical decoupling sequences that made \emeab{} successful. This raises an interesting question about the extent of entanglement enhancement in learning quantum processes in the presence of some locality constraints.

Additionally, since SPAM errors contribute to a major part of the overhead, techniques such as heralded state preparation can be used to improve the error mitigation overhead~\cite{heraldedstateprep}. Moreover, assessing this overhead in different quantum computing platforms, especially those with flexible qubit layout such as trapped ions~\cite{pino2021demonstration} and neutral atoms~\cite{bluvstein2024logical} where qubit can be shuttled, may reveal interesting trade-offs between the connectivity, speed, and quality of the platforms. 

In the future,
as we move towards realizing fault-tolerant quantum memory, it would be interesting to examine applications of (partially) error-corrected memory in learning properties of other noisy physical systems, thus bridging the gap between error-mitigated learning and error-corrected quantum metrology~\cite{zhou2018achieving}.

\section{Methods}

\subsection{Notations and Basics}

We follow the notations of Ref.~\cite{chen2022quantum}.
Let $\Pn=\{ I, X, Y, Z\}^{\otimes n}$ be the Pauli group modulo the global phase. The Abelian group $\Pn$ is isomorphic to $\mbb Z^{2n}_2$.
In other words, each element of $\Pn$ uniquely corresponds to a $2n$-bit string
\begin{equation}\label{eq:pauli_string}
    a = (a_x,a_z) = a_{x,1}\cdots a_{x,n}a_{z,1}\cdots a_{z,n},
\end{equation}
where $a_x\coleq a_{x,1}\cdots a_{x_n}$ and $a_z\coleq a_{z,1}\cdots a_{z_n}$ are elements of $\mbb Z_2^n$.
The corresponding Pauli operator is given by
\begin{equation}    
	P_a = \bigotimes_{j=1}^n i^{a_{x,j}a_{z,j}}  X^{a_{x,j}} Z^{a_{z,j}},
\end{equation}
where the phase is chosen to ensure Hermiticity.
Besides the standard addition and inner product defined over $\mbb Z^{2n}_{2}$, we define the sympletic inner product over $\Pn$ as
\begin{equation}
	\expval{a,b} = \sum_{j=1}^n (a_{x,j}b_{z,j}+a_{z,j}b_{x,j}) \mod{2}.
\end{equation}
One can verify that $\expval{a,b}=0$ if $P_a$ and $P_b$ commute; otherwise, $\expval{a,b}=1$.
In the following, we use the label $a$ and the Pauli operator $P_a$ interchangeably when there is no confusion.
Finally, we define the \emph{Pauli pattern} of $P_a$ as $\pt(a)$, where $\pt:\Pn\mapsto\mbb Z^n_2$ maps the $i$th Pauli operator to $0$ if it is $I$ and $1$ otherwise. For example, $\pt(XYIZI)=11010$.
The Hamming weight of $\pt(a)$ is simply called the weight of $P_a$.

\medskip
\noindent An $n$-qubit Pauli channel $\Lambda$ is a quantum channel of the following form
\begin{equation}
	\Lambda(\cdot) = \sum_{a\in\Pn}p_a P_a(\cdot)P_a,
\end{equation}
where $\{p_a\}_a$ are the \emph{Pauli error rates} that form a $4^n$-dimensional probability distribution.
An alternative expression for $\Lambda$ is
\begin{equation}
	\Lambda(\cdot) = \frac{1}{2^n}\sum_{b\in\Pn}\lambda_b\Tr(P_b(\cdot))P_b,
\end{equation}
where $\{\lambda_b\}_b$ are the \emph{Pauli eigenvalues} or \emph{Pauli fidelities}. 
These two sets of parameters are related by the Walsh-Hadamard transform~\cite{flammia2020efficient},
\small{
\begin{equation}
	\begin{aligned}
		\lambda_b = \sum_{a\in\Pn}p_a(-1)^{\expval{a,b}},\quad
		p_a = \frac{1}{4^n}\sum_{b\in\Pn}\lambda_b(-1)^{\expval{a,b}}.
	\end{aligned}
\end{equation}}
\noindent We assume $\lambda_b>0$ for all $b\in\Pn$, which should be satisfied by any reasonable gate noise channel $\Lambda$.

\medskip

Let $\{\ket{j}\}_{j=0}^{2^n-1}$ denote the $n$-qubit computational basis. We call $\ket{\Phi}\coleq \sum_{j=0}^{2^n-1}\ket{jj}/\sqrt{2^n}$ the (canonical) Bell state. The Bell basis $\{\ket\Phi_\nu\}_{\nu\in\Pn}$ is defined as $\ket{\Phi}_\nu\coleq  I\otimes P_\nu\ket{\Phi}$, which forms a orthonormal basis for the $2n$-qubit Hilbert space.
\subsection{Details of \emeab{} protocol}

We focus on characterizing a noisy $n$-qubit Clifford gate $\widetilde{\mc G}$.
Suppose the gate noise is Markovian and time-stationary, in which case we can model $\widetilde{\mc G} = \mc G\circ\Lambda$, where $\mc G$ is the ideal noise-free gate and $\Lambda$ is the noise channel. Let $\Lambda^{\sf P}$ be the Pauli twirled version of $\Lambda$. The ideal goal is to learn the Pauli fidelities $\{\lambda_a\}_{a\in\Pn}$ of $\Lambda^{\sf P}$ in a SPAM-robust manner. 
However, it has been proven that certain Pauli fidelities cannot be identified independently from the noisy SPAM~\cite{chen2023learnability}. To circumvent this issue, we introduce the $\mc G$-symmetrized Pauli fidelity as
\begin{equation}
    \bar\lambda_a\coleq \left(\prod_{k=0}^{d_0-1} \lambda_{\mc G^k(a)} \right)^{1/d_0},\quad \forall a\in\Pn.
\end{equation}
Here $d_0$ is the smallest positive integer such that $\mc G^{d_0} = {\mc I}$. $\lambda_{\mc G^k(a)}$ denotes the fidelity corresponding to $\mc G^k(P_a)$ with a potential $\pm$ sign ignored. 
As an example, the symmetrized Pauli eigenvalues for a single CNOT are listed in Table~\ref{tab:CNOT_sym}. 
\setlength{\tabcolsep}{16pt}
\begin{table}[!htp]
    \centering
    \begin{tabular}{l l}
       \toprule
       $a$ & $\bar{\lambda}_a$\\
       \midrule
       ZI & $\lambda_{ZI}$ \\
       IX & $\lambda_{IX}$ \\
       ZX & $\lambda_{ZX}$ \\
       IZ,ZZ & $(\lambda_{IZ}\lambda_{ZZ})^{1/2}$ \\
       XI,XX & $(\lambda_{XI}\lambda_{XX})^{1/2}$ \\
       XZ,YY & $(\lambda_{XZ}\lambda_{YY})^{1/2}$ \\
       YI,YX & $(\lambda_{YI}\lambda_{YX})^{1/2}$ \\
       IY,ZY & $(\lambda_{IZ}\lambda_{ZY})^{1/2}$ \\
       XY,YZ & $(\lambda_{XY}\lambda_{YZ})^{1/2}$\\
       \bottomrule
    \end{tabular}
    \caption{Symmetrized Pauli eigenvalues for a single CNOT gate.}
    \label{tab:CNOT_sym}
\end{table}
The symmetrized Pauli fidelity $\bar\lambda_a$ are SPAM-robustly learnable and has been reported in the literature~\cite{erhard2019characterizing,chen2023learnability,carignan2023error,hashim2020randomized}, thus our goal will be learning these quantities.
Since \emeab{} involves an auxiliary quantum memory, we recall our assumptions that the gate noise has no correlation with the ancilla noise, i.e., 
\begin{equation}
    \widetilde{{\mc I}\otimes\mc G} = ({\mc I}\otimes\mc G)\circ(\Lambda'\otimes\Lambda).
\end{equation}
We do allow generic SPAM noise between the system and ancilla for the Bell state preparation and Bell measurements.
Finally, we make the standard assumption that single-qubit gates have negligible noise (or absorbed into the entangling gate layer).

\begin{figure}[!htp]
\begin{algorithm}[H]
	\caption{\emeab{}}
	\label{alg:emeab}
	\begin{algorithmic}[1]
		\Require $N_C,N_S,\bm D, N_C',N_S',\bm D',\bm Q.$

        \State $\bm \mu^\eab{} = \Call{{Collect}\eab{}}{N_C,N_S,\bm D}$
        \State $\bm k^\auxb{} = \Call{{Collect}\auxb{}}{N_C',N_S',\bm D'}$
		\For{$a\in \bm Q$}
		\State $\hat\lambda_{\eab{},a} \coleq \Call{{Fit}\eab{}}{\bm \mu^\eab{},a}$
  		\State $\hat\lambda_{\auxb{},\pt(a)} \coleq \Call{{Fit}\auxb{}}{\bm k^\auxb{},\pt(a)}$
            \State $\hat\lambda_{a} \coleq \hat\lambda_{\eab{},a}/\hat\lambda_{\auxb{},\pt(a)}$
		\EndFor
		\State \Return{$\{\hat{\lambda}_{a}\}_{a\in\bm Q}$}.
	\end{algorithmic}
\end{algorithm}
\end{figure}

The procedure of \emeab{} is described in Algorithm~\ref{alg:emeab}. In the input parameters, $\bm D = [d_1,\cdots,d_{|\bm D|}]$ is a list of circuit depths. We require each $d_i$ to be an integer multiple of $d_0$. See the next section for how to choose $\bm D$.  $N_C$, $N_S$ are number of random circuits per depth and number of measurement shots per circuit, respectively. The unprimed, primed version are parameters for $\eab{}$ and $\auxb{}$, respectively; $\bm Q = [a_1,\cdots, a_{|\bm Q|}]\subseteq\Pn$ is set of queries of Pauli fidelities. The protocol will return an estimator $\hat\lambda_a$ to the symmetrized Pauli fidelity $\bar\lambda_a$ for each $a\in\bm Q$. 

\begin{figure}[!htp]
\begin{algorithm}[H]
\begin{algorithmic}[1]
    \caption{Subroutines for \emeab{}}
    \label{alg:sub}
    \Procedure{SampleGateSequence}{$d$}
        \State Uniformly sample $d$ layers of single-qubit Clifford gates $C_1,\cdots,C_d$ and $d+2$ Pauli gates $P_{a_1},\cdots,P_{a_d},P_\alpha,P_\beta$.
        \State $C_{\mr{end}}\coleq P_\alpha C_1^\dagger C_2^\dagger \cdots C_d^\dagger$.
        \State $P_{\mr{end}}\coleq P_\beta \mc G^d(P_{a_1})\mc G^{d-1}(P_{a_2})\cdots\mc G(P_{a_d})$.
        \State Define $\mc T$ to be the following gate sequence
        
        \begin{quantikz}[column sep=5.5pt, row sep={20pt,between origins}]
            & \gate{C_\mr{1}}  & \gate{{\mc I}} &  \gate{C_\mr{2}}       &\qw&\cdots~&\gate{C_d} & \gate{{\mc I}}  & \gate{C_\mr{end}}   & \qw \\
            & \gate[style={inner xsep=-1.7pt}]{P_{a_1}}  &  \gate{\mc{G}} &  \gate[style={inner xsep=-1.7pt}]{P_{a_2}}     &\qw&\cdots~&\gate[style={inner xsep=-1.7pt}]{P_{a_d}} & \gate{\mc G}  & \gate{P_\mr{end}}   & \qw
        \end{quantikz}
    \State \Return{$\mc T,\alpha,\beta$}
    \EndProcedure
    \\\hrulefill
    \Procedure{Collect\eab{}}{$N_C,N_S,\bm D$}
    \For{$d\in\bm D$}
    \For{$i=1\cdots N_C$}
    \State $\mc T,\alpha,\beta\coleq \Call{SampleGateSequence}{d}$
    \For{$j=1\cdots N_S$}
    \State Prepare $\ket{\Phi}$, apply $\mc T$, measure in $\{\ket{\Phi_\nu}\}$. Denote outcome as $\nu$.
    \State $\mu_{d,i,j} \coleq \nu + \alpha + \beta \in \Pn\cong\mbb Z_2^{2n}$.
    \EndFor
    \EndFor
    \EndFor
    \State \Return{$\bm \mu\coleq\{\mu_{d,i,j}\}$}
    \EndProcedure
    \\\hrulefill
    \Procedure{Collect\auxb{}}{$N_C,N_S,\bm D$}
    \For{$d\in\bm D$}
    \For{$i=1\cdots N_C$}
    \State $\mc T,\alpha,\beta\coleq \Call{SampleGateSequence}{d}$
    \For{$j=1\cdots N_S$}
    \State Prepare $\ket{0}^{\otimes 2n}$, apply $\mc T$, measure the ancilla in $\{{\ket l}\}$. Denote outcome as $l$.
    \State $k_{d,i,j} \coleq l + \alpha_x\in\mbb Z^n_2$.
    \EndFor
    \EndFor
    \EndFor
    \State \Return{$\bm k\coleq\{k_{d,i,j}\}$}
    \EndProcedure
    \\\hrulefill
    \Procedure{Fit\eab{}}{$\bm \mu,a$}
    \State Obtain $N_C,N_S,\bm D$ from $\bm\mu$.
    \For{$d\in\bm D$}
        \State $\hat f_d\coleq \frac{1}{N_CN_S}\sum_{i,j}(-1)^{\expval{\mu,a}}$
    \EndFor
    \State Fit $\{\hat f_d\}$ to $f_d = \hat\alpha_a^\eab{}(\hat\lambda_a^{\eab{}})^d$.
    \State \Return{$\hat{\lambda}_a^{\eab{}}$}
    \EndProcedure
    \\\hrulefill
    \Procedure{Fit\auxb{}}{$\bm k,s$}
    \State Obtain $N_C,N_S,\bm D$ from $\bm k$.
    \For{$d\in\bm D$}
        \State $\hat f_d\coleq \frac{1}{N_CN_S}\sum_{i,j}(-1)^{k\cdot s}$
    \EndFor
    \State Fit $\{\hat f_d\}$ to $f_d = \hat\alpha_s^\auxb{}(\hat\lambda_s^{\auxb{}})^d$.
    \State \Return{$\hat{\lambda}_s^{\auxb{}}$}
    \EndProcedure    
\end{algorithmic}
\end{algorithm}
\end{figure}

The subroutines used for \emeab{} are described in Algorithm~\ref{alg:sub}. \textsc{SampleGateSequence} generates a randomized gate sequence containing $d$ layers of $\mc G$. Here $\{P_{a_i}\}$ are the Pauli twirling gates while $\{C_i\}$ are the local-Clifford twirling gates.
The $P_\alpha$ and $P_\beta$ are Pauli gates for measurement twirling.
The identity gate will be experimentally implemented as a dynamical decoupling sequence that suppresses idling noise and system-ancilla crosstalk~\cite{seif2024suppressing}.
\textsc{Collect\eab{}} calls \textsc{SampleGateSequence} as a subroutine to construct {\eab{}} experiments. Note that, $\mu_{d,i,j}$ defined in Line 14 is the ``twirling-corrected'' measurement outcome taking into account $P_\alpha,P_\beta$. Similar for \textsc{Collect\auxb{}}. The $\alpha_x$ in Line 23 is the ``X part'' of the Pauli operator as defined in Eq.~\eqref{eq:pauli_string}.
Finally, \textsc{Fit\eab{}} and \textsc{Fit\auxb{}} extract the fidelity per layer via curve fitting to cancel the SPAM noise, as is standard for RB-type methods~\cite{knill2008randomized,magesan2011scalable,flammia2020efficient}.

In SM Sec. III, we prove that for any $a$, $\hat\lambda_a$ given by \emeab{} is a consistent (\textit{i.e.}, asymptotically unbiased) estimator for $\bar\lambda_a$. For this purpose, we will show that $\hat\lambda_{\eab{},a}$ is a consistent estimator for $\bar\lambda_a\lambda'_{\pt(a)}$, where $\lambda'_a$ denotes Pauli fidelities of the ancilla noise channel $\Lambda'$, and $\hat\lambda_{\auxb{},a}$ is a consistent estimator for $\lambda'_{\pt(a)}$. Taking ratio of these two thus yields a consistent estimator for $\bar\lambda_a$.

\subsection{Resource cost analysis}

In this section, we analyze the variance overhead of \emeab{} in estimating individual symmetrized Pauli eigenvalues, and compare it with the experimental setting reduction.
Consider a specific $a\in\Pn$ with weight $w$. Henceforth, we omit the subscript of $a$ or $\pt(a)$ for notational conciseness. 
Since $\hat\lambda_\emeab\coleq \hat\lambda_\eab/\hat\lambda_\auxb$, assuming both $\hat\lambda_\eab{}$ and $\hat\lambda_\auxb{}$ are sufficiently close to the true values, using the first-order Taylor expansion yields,
\small{
\begin{equation}\label{eq:lin_variance}
    \frac{\mr{Var}[\hat\lambda_\emeab]}{\mr{Var}[\hat\lambda_{\sysb}]} \approx 
    \frac{1}{\lambda_\auxb^2}\frac{\mr{Var}[\hat\lambda_\eab]}{\mr{Var}[\hat\lambda_\sysb]} + \frac{\lambda_\sysb^2}{\lambda_\auxb^2}\frac{\mr{Var}[\hat\lambda_\auxb]}{\mr{Var}[\hat\lambda_\sysb]}.
\end{equation}
}
The second term uses the assumption that $\lambda_\eab=\lambda_\sysb\lambda_\auxb$, i.e., the system and auxiliary noise are independent.
Note that $\hat\lambda_{\eab}$, $\hat\lambda_{\auxb}$, $\hat\lambda_{\sysb}$ are all obtained by fitting an exponential decay in the form of $f_{d,k} = \alpha \lambda^d$.
Omitting the label of protocols for now.
For simplicity, we assume each protocol only uses two circuit depths, $\{0,d\}$, in which case the fidelity estimator is given by $\hat\lambda\coleq (\hat f_d/\hat f_0)^{1/d}$, called a ratio estimator. 
Indeed, let $\hat f_{0(d)} \coleq f_{0(d)} + \hat\varepsilon_{0(d)}$,
\begin{equation}
\begin{aligned}           
    \hat\lambda &= \left(\frac{\alpha\lambda^d + \hat \varepsilon_d}{\alpha + \hat \varepsilon_0}\right)^{\frac1d}
    \approx \lambda \left(1 + \frac{\hat\varepsilon_d -\lambda^d\hat\varepsilon_0}{d\alpha\lambda^d}\right),
\end{aligned}
\end{equation}
up to first-order approximation in $\hat\varepsilon_{0(d)}$. Therefore, the variance of $\hat\lambda$ is given by
\begin{equation}
    \mr{Var}[\hat\lambda] \approx \frac{\lambda^2}{d^2\alpha^2}\left(\mr{Var}[\hat f_0] + \mr{Var}[\hat f_d]/\lambda^{2d}\right) \eqcol \frac{\lambda^2V}{d^2\alpha^2},
\end{equation}
where $V$ is a short-hand notation for the term within the bracket.
Putting this back into Eq.~\eqref{eq:lin_variance} yields,
\begin{equation}\label{eq:up_to_here_works}
    \frac{\mr{Var}[\hat\lambda_\emeab]}{\mr{Var}[\hat\lambda_{\sysb}]}\approx 
    \frac{d_\sysb^2\alpha_\sysb^2V_\eab}{d_\eab^2\alpha_\eab^2V_\sysb} + \frac{d_\sysb^2\alpha_\sysb^2V_\auxb}{d_\auxb^2\alpha_\auxb^2V_\sysb}.
\end{equation}
To proceed, we need to decide $d$ for each protocols. One sensible choice is to let $\lambda^d\approx 1/e$, which enables a multiplicative precision estimation for $\lambda$ with respect to the infidelity. There exists a procedure to efficiently search for such $d$~\cite{harper2019statistical}. Here we assume the search is done \textit{a priori}.
On the other hand, if we assume that (1) $f_0$ is sufficiently close to $1$, so the contribution of $\mr{Var}[\hat f_0]$ to $V$ can be ignored, and that (2) $\hat f_d$ has binomial variance $\mr{Var}[\hat f_d] = (1-f_d)^2/N$ (which is true if $N_S=1$ or the noise is sufficiently uniform among different random circuit instances), then the value of $V$ is comparable for all protocols. 
Finally, if we further make the assumption that $\lambda_\auxb{}$ is at least as large as $\lambda_\sysb{}$ (which means the system gate noise is at least as large as the ancilla idling noise), we have
\begin{equation}\label{eq:final_overhead}
\begin{aligned}
    \frac{\mr{Var}[\hat\lambda_\emeab]}{\mr{Var}[\hat\lambda_{\sysb}]}&\approx
    \left(1+\frac{r_\auxb}{r_\sysb{}}\right)^2\frac{\alpha_\sysb^2}{\alpha^2_\eab{}} + 1.
\end{aligned}
\end{equation}
Here $r\coleq\ln\lambda^{-1}\approx 1-\lambda$ when $\lambda$ is close to $1$. In conclusion, under all the assumptions we made in above, the variance overhead of \emeab{} roughly scales as the SPAM fidelity ratio $(\alpha_\sysb/\alpha_\eab)^2$. 
Suppose each Bell pair suffers from identical and independent error, in which case $\alpha_\eab{} = c^{w/2}$ for some constant $c<1$. As long as $c>1/2$ (which is a mild assumption about the Bell pair fidelity), the variance overhead will grow slower than $2^w$, and thus \emeab{} will have an efficiency enhancement over \sysb{}.

We remark that the above analysis only aims at understanding the rough scaling of the variance overhead rather than giving an exact prediction. In our experiments, not all assumptions made above are satisfied: For example, we choose $d=0,16,32$ and extract the rate using a fit for all protocols instead of searching for individual optimal depths and using the ratio estimator. We also choose $N_S=500$, therefore $\mr{Var}[\hat 
f_d]$ does not satisfy binomial variance. Nevertheless, by directly estimating the sample variance, we obtain a variance overhead scales as $1.33^w$ in  Fig.~\ref{fig:multieab}, which is considerably smaller than $2^w$ or $3^w$, demonstrating the entanglement enhancement even in presence of noise. We also verify that, although Eq.~\eqref{eq:final_overhead} deviates from the empirical variance overhead, Eq.~\eqref{eq:up_to_here_works} gives a precise prediction by inserting the empirical variance for each $\mr{Var}[\hat f_{0}]$ and $\mr{Var}[\hat f_{d}]$ at $d=32$. See SM Sec. IV.

\begin{acknowledgements}
We thank Elisa Bäumer for performing experiments at the beginning of this project. We thank Joseph Emerson, Luke Govia, Hsin-Yuan Huang, Abhinav Kandala, Youngseok Kim, Yunchao Liu,   David McKay, Kristan Temme, Ming Yuan for helpful discussions.
{This research was supported in part by grant NSF PHY-2309135 to the Kavli Institute for Theoretical Physics (KITP).}
S.C. \& L.J. acknowledge support from the ARO (W911NF-23-1-0077), ARO MURI (W911NF-21-1-0325), AFOSR MURI (FA9550-19-1-0399, FA9550-21-1-0209, FA9550-23-1-0338), NSF (OMA-1936118, ERC-1941583, OMA-2137642, OSI-2326767, CCF-2312755), NTT Research, Packard Foundation (2020-71479). A.S. and S.M. were sponsored by the Army Research Office  under Grant Number W911NF-21-1-0002. The views and conclusions contained in this document are those of the authors and should not be interpreted as representing the official policies, either expressed or implied, of the Army Research Office or the U.S. Government. The U.S. Government is authorized to reproduce and distribute reprints for Government purposes notwithstanding any copyright notation herein.
\end{acknowledgements}
\bibliographystyle{naturemag}
\bibliography{paulilearning}

\begin{thebibliography}{10}
\expandafter\ifx\csname url\endcsname\relax
  \def\url#1{\texttt{#1}}\fi
\expandafter\ifx\csname urlprefix\endcsname\relax\def\urlprefix{URL }\fi
\providecommand{\bibinfo}[2]{#2}
\providecommand{\eprint}[2][]{\url{#2}}

\bibitem{temme2017error}
\bibinfo{author}{Temme, K.}, \bibinfo{author}{Bravyi, S.} \&
  \bibinfo{author}{Gambetta, J.~M.}
\newblock \bibinfo{title}{Error mitigation for short-depth quantum circuits}.
\newblock \emph{\bibinfo{journal}{Physical review letters}}
  \textbf{\bibinfo{volume}{119}}, \bibinfo{pages}{180509}
  (\bibinfo{year}{2017}).

\bibitem{van2023probabilistic}
\bibinfo{author}{van Den~Berg, E.}, \bibinfo{author}{Minev, Z.~K.},
  \bibinfo{author}{Kandala, A.} \& \bibinfo{author}{Temme, K.}
\newblock \bibinfo{title}{Probabilistic error cancellation with sparse
  pauli--lindblad models on noisy quantum processors}.
\newblock \emph{\bibinfo{journal}{Nature Physics}} \bibinfo{pages}{1--6}
  (\bibinfo{year}{2023}).

\bibitem{kim2023evidence}
\bibinfo{author}{Kim, Y.} \emph{et~al.}
\newblock \bibinfo{title}{Evidence for the utility of quantum computing before
  fault tolerance}.
\newblock \emph{\bibinfo{journal}{Nature}} \textbf{\bibinfo{volume}{618}},
  \bibinfo{pages}{500--505} (\bibinfo{year}{2023}).

\bibitem{lidar2013quantum}
\bibinfo{author}{Lidar, D.~A.} \& \bibinfo{author}{Brun, T.~A.}
\newblock \emph{\bibinfo{title}{Quantum error correction}}
  (\bibinfo{publisher}{Cambridge university press},
  \bibinfo{address}{Cambridge}, \bibinfo{year}{2013}).

\bibitem{RevModPhys.89.035002}
\bibinfo{author}{Degen, C.~L.}, \bibinfo{author}{Reinhard, F.} \&
  \bibinfo{author}{Cappellaro, P.}
\newblock \bibinfo{title}{Quantum sensing}.
\newblock \emph{\bibinfo{journal}{Rev. Mod. Phys.}}
  \textbf{\bibinfo{volume}{89}}, \bibinfo{pages}{035002}
  (\bibinfo{year}{2017}).
\newblock
  \urlprefix\url{https://link.aps.org/doi/10.1103/RevModPhys.89.035002}.

\bibitem{PhysRevLett.96.010401}
\bibinfo{author}{Giovannetti, V.}, \bibinfo{author}{Lloyd, S.} \&
  \bibinfo{author}{Maccone, L.}
\newblock \bibinfo{title}{Quantum metrology}.
\newblock \emph{\bibinfo{journal}{Phys. Rev. Lett.}}
  \textbf{\bibinfo{volume}{96}}, \bibinfo{pages}{010401}
  (\bibinfo{year}{2006}).
\newblock
  \urlprefix\url{https://link.aps.org/doi/10.1103/PhysRevLett.96.010401}.

\bibitem{PhysRevLett.90.193601}
\bibinfo{author}{Altepeter, J.~B.} \emph{et~al.}
\newblock \bibinfo{title}{Ancilla-assisted quantum process tomography}.
\newblock \emph{\bibinfo{journal}{Phys. Rev. Lett.}}
  \textbf{\bibinfo{volume}{90}}, \bibinfo{pages}{193601}
  (\bibinfo{year}{2003}).
\newblock
  \urlprefix\url{https://link.aps.org/doi/10.1103/PhysRevLett.90.193601}.

\bibitem{mohseni2008quantum}
\bibinfo{author}{Mohseni, M.}, \bibinfo{author}{Rezakhani, A.~T.} \&
  \bibinfo{author}{Lidar, D.~A.}
\newblock \bibinfo{title}{Quantum-process tomography: Resource analysis of
  different strategies}.
\newblock \emph{\bibinfo{journal}{Physical Review A}}
  \textbf{\bibinfo{volume}{77}}, \bibinfo{pages}{032322}
  (\bibinfo{year}{2008}).

\bibitem{PhysRevLett.129.133601}
\bibinfo{author}{Xue, S.} \emph{et~al.}
\newblock \bibinfo{title}{Variational entanglement-assisted quantum process
  tomography with arbitrary ancillary qubits}.
\newblock \emph{\bibinfo{journal}{Phys. Rev. Lett.}}
  \textbf{\bibinfo{volume}{129}}, \bibinfo{pages}{133601}
  (\bibinfo{year}{2022}).
\newblock
  \urlprefix\url{https://link.aps.org/doi/10.1103/PhysRevLett.129.133601}.

\bibitem{PhysRevLett.86.4195}
\bibinfo{author}{D'Ariano, G.~M.} \& \bibinfo{author}{Lo~Presti, P.}
\newblock \bibinfo{title}{Quantum tomography for measuring experimentally the
  matrix elements of an arbitrary quantum operation}.
\newblock \emph{\bibinfo{journal}{Phys. Rev. Lett.}}
  \textbf{\bibinfo{volume}{86}}, \bibinfo{pages}{4195--4198}
  (\bibinfo{year}{2001}).
\newblock \urlprefix\url{https://link.aps.org/doi/10.1103/PhysRevLett.86.4195}.

\bibitem{aharonov2022quantum}
\bibinfo{author}{Aharonov, D.}, \bibinfo{author}{Cotler, J.} \&
  \bibinfo{author}{Qi, X.-L.}
\newblock \bibinfo{title}{Quantum algorithmic measurement}.
\newblock \emph{\bibinfo{journal}{Nature communications}}
  \textbf{\bibinfo{volume}{13}}, \bibinfo{pages}{887} (\bibinfo{year}{2022}).

\bibitem{chen2022quantum}
\bibinfo{author}{Chen, S.}, \bibinfo{author}{Zhou, S.}, \bibinfo{author}{Seif,
  A.} \& \bibinfo{author}{Jiang, L.}
\newblock \bibinfo{title}{Quantum advantages for pauli channel estimation}.
\newblock \emph{\bibinfo{journal}{Physical Review A}}
  \textbf{\bibinfo{volume}{105}}, \bibinfo{pages}{032435}
  (\bibinfo{year}{2022}).

\bibitem{chen2022exponential}
\bibinfo{author}{Chen, S.}, \bibinfo{author}{Cotler, J.},
  \bibinfo{author}{Huang, H.-Y.} \& \bibinfo{author}{Li, J.}
\newblock \bibinfo{title}{Exponential separations between learning with and
  without quantum memory}.
\newblock In \emph{\bibinfo{booktitle}{2021 IEEE 62nd Annual Symposium on
  Foundations of Computer Science (FOCS)}}, \bibinfo{pages}{574--585}
  (\bibinfo{organization}{IEEE}, \bibinfo{year}{2022}).

\bibitem{huang2022quantum}
\bibinfo{author}{Huang, H.-Y.} \emph{et~al.}
\newblock \bibinfo{title}{Quantum advantage in learning from experiments}.
\newblock \emph{\bibinfo{journal}{Science}} \textbf{\bibinfo{volume}{376}},
  \bibinfo{pages}{1182--1186} (\bibinfo{year}{2022}).

\bibitem{caro2022learning}
\bibinfo{author}{Caro, M.~C.}
\newblock \bibinfo{title}{Learning quantum processes and hamiltonians via the
  pauli transfer matrix}.
\newblock \emph{\bibinfo{journal}{ACM Transactions on Quantum Computing}}
  \textbf{\bibinfo{volume}{5}}, \bibinfo{pages}{1--53} (\bibinfo{year}{2024}).

\bibitem{PhysRevLett.79.3865}
\bibinfo{author}{Huelga, S.~F.} \emph{et~al.}
\newblock \bibinfo{title}{Improvement of frequency standards with quantum
  entanglement}.
\newblock \emph{\bibinfo{journal}{Phys. Rev. Lett.}}
  \textbf{\bibinfo{volume}{79}}, \bibinfo{pages}{3865--3868}
  (\bibinfo{year}{1997}).
\newblock \urlprefix\url{https://link.aps.org/doi/10.1103/PhysRevLett.79.3865}.

\bibitem{escher2011general}
\bibinfo{author}{Escher, B.}, \bibinfo{author}{de~Matos~Filho, R.~L.} \&
  \bibinfo{author}{Davidovich, L.}
\newblock \bibinfo{title}{General framework for estimating the ultimate
  precision limit in noisy quantum-enhanced metrology}.
\newblock \emph{\bibinfo{journal}{Nature Physics}}
  \textbf{\bibinfo{volume}{7}}, \bibinfo{pages}{406--411}
  (\bibinfo{year}{2011}).

\bibitem{demkowicz2012elusive}
\bibinfo{author}{Demkowicz-Dobrza{\'n}ski, R.},
  \bibinfo{author}{Ko{\l}ody{\'n}ski, J.} \& \bibinfo{author}{Gu{\c{t}}{\u{a}},
  M.}
\newblock \bibinfo{title}{The elusive heisenberg limit in quantum-enhanced
  metrology}.
\newblock \emph{\bibinfo{journal}{Nature communications}}
  \textbf{\bibinfo{volume}{3}}, \bibinfo{pages}{1063} (\bibinfo{year}{2012}).

\bibitem{harper2020efficient}
\bibinfo{author}{Harper, R.}, \bibinfo{author}{Flammia, S.~T.} \&
  \bibinfo{author}{Wallman, J.~J.}
\newblock \bibinfo{title}{Efficient learning of quantum noise}.
\newblock \emph{\bibinfo{journal}{Nature Physics}}
  \textbf{\bibinfo{volume}{16}}, \bibinfo{pages}{1184--1188}
  (\bibinfo{year}{2020}).

\bibitem{erhard2019characterizing}
\bibinfo{author}{Erhard, A.} \emph{et~al.}
\newblock \bibinfo{title}{Characterizing large-scale quantum computers via
  cycle benchmarking}.
\newblock \emph{\bibinfo{journal}{Nature communications}}
  \textbf{\bibinfo{volume}{10}}, \bibinfo{pages}{5347} (\bibinfo{year}{2019}).

\bibitem{ferracin2022efficiently}
\bibinfo{author}{Ferracin, S.} \emph{et~al.}
\newblock \bibinfo{title}{Efficiently improving the performance of noisy
  quantum computers}.
\newblock \emph{\bibinfo{journal}{arXiv preprint arXiv:2201.10672}}
  (\bibinfo{year}{2022}).

\bibitem{aliferis2008fault}
\bibinfo{author}{Aliferis, P.} \& \bibinfo{author}{Preskill, J.}
\newblock \bibinfo{title}{Fault-tolerant quantum computation against biased
  noise}.
\newblock \emph{\bibinfo{journal}{Physical Review A}}
  \textbf{\bibinfo{volume}{78}}, \bibinfo{pages}{052331}
  (\bibinfo{year}{2008}).

\bibitem{chubb2021statistical}
\bibinfo{author}{Chubb, C.~T.} \& \bibinfo{author}{Flammia, S.~T.}
\newblock \bibinfo{title}{Statistical mechanical models for quantum codes with
  correlated noise}.
\newblock \emph{\bibinfo{journal}{Annales de l’Institut Henri Poincar{\'e}
  D}} \textbf{\bibinfo{volume}{8}}, \bibinfo{pages}{269--321}
  (\bibinfo{year}{2021}).

\bibitem{chen2024tight}
\bibinfo{author}{Chen, S.}, \bibinfo{author}{Oh, C.}, \bibinfo{author}{Zhou,
  S.}, \bibinfo{author}{Huang, H.-Y.} \& \bibinfo{author}{Jiang, L.}
\newblock \bibinfo{title}{Tight bounds on pauli channel learning without
  entanglement}.
\newblock \emph{\bibinfo{journal}{Physical Review Letters}}
  \textbf{\bibinfo{volume}{132}}, \bibinfo{pages}{180805}
  (\bibinfo{year}{2024}).

\bibitem{chen2023futility}
\bibinfo{author}{Chen, S.} \& \bibinfo{author}{Gong, W.}
\newblock \bibinfo{title}{Futility and utility of a few ancillas for pauli
  channel learning}.
\newblock \emph{\bibinfo{journal}{arXiv preprint arXiv:2309.14326}}
  (\bibinfo{year}{2023}).

\bibitem{fawzi2023lower}
\bibinfo{author}{Fawzi, O.}, \bibinfo{author}{Oufkir, A.} \&
  \bibinfo{author}{Fran{\c{c}}a, D.~S.}
\newblock \bibinfo{title}{Lower bounds on learning pauli channels}.
\newblock \emph{\bibinfo{journal}{arXiv preprint arXiv:2301.09192}}
  (\bibinfo{year}{2023}).

\bibitem{nielsen2010quantum}
\bibinfo{author}{Nielsen, M.~A.} \& \bibinfo{author}{Chuang, I.~L.}
\newblock \emph{\bibinfo{title}{Quantum computation and quantum information}}
  (\bibinfo{publisher}{Cambridge university press}, \bibinfo{year}{2010}).

\bibitem{flammia2020efficient}
\bibinfo{author}{Flammia, S.~T.} \& \bibinfo{author}{Wallman, J.~J.}
\newblock \bibinfo{title}{Efficient estimation of pauli channels}.
\newblock \emph{\bibinfo{journal}{ACM Transactions on Quantum Computing}}
  \textbf{\bibinfo{volume}{1}}, \bibinfo{pages}{1--32} (\bibinfo{year}{2020}).

\bibitem{wallman2016noise}
\bibinfo{author}{Wallman, J.~J.} \& \bibinfo{author}{Emerson, J.}
\newblock \bibinfo{title}{Noise tailoring for scalable quantum computation via
  randomized compiling}.
\newblock \emph{\bibinfo{journal}{Physical Review A}}
  \textbf{\bibinfo{volume}{94}}, \bibinfo{pages}{052325}
  (\bibinfo{year}{2016}).

\bibitem{hashim2020randomized}
\bibinfo{author}{Hashim, A.} \emph{et~al.}
\newblock \bibinfo{title}{Randomized compiling for scalable quantum computing
  on a noisy superconducting quantum processor}.
\newblock \emph{\bibinfo{journal}{arXiv preprint arXiv:2010.00215}}
  (\bibinfo{year}{2020}).

\bibitem{ware2021experimental}
\bibinfo{author}{Ware, M.} \emph{et~al.}
\newblock \bibinfo{title}{Experimental pauli-frame randomization on a
  superconducting qubit}.
\newblock \emph{\bibinfo{journal}{Phys. Rev. A}}
  \textbf{\bibinfo{volume}{103}}, \bibinfo{pages}{042604}
  (\bibinfo{year}{2021}).
\newblock \urlprefix\url{https://link.aps.org/doi/10.1103/PhysRevA.103.042604}.

\bibitem{choi1975completely}
\bibinfo{author}{Choi, M.-D.}
\newblock \bibinfo{title}{Completely positive linear maps on complex matrices}.
\newblock \emph{\bibinfo{journal}{Linear algebra and its applications}}
  \textbf{\bibinfo{volume}{10}}, \bibinfo{pages}{285--290}
  (\bibinfo{year}{1975}).

\bibitem{bennett1992communication}
\bibinfo{author}{Bennett, C.~H.} \& \bibinfo{author}{Wiesner, S.~J.}
\newblock \bibinfo{title}{Communication via one-and two-particle operators on
  einstein-podolsky-rosen states}.
\newblock \emph{\bibinfo{journal}{Physical review letters}}
  \textbf{\bibinfo{volume}{69}}, \bibinfo{pages}{2881} (\bibinfo{year}{1992}).

\bibitem{shedon2016cr}
\bibinfo{author}{Sheldon, S.}, \bibinfo{author}{Magesan, E.},
  \bibinfo{author}{Chow, J.~M.} \& \bibinfo{author}{Gambetta, J.~M.}
\newblock \bibinfo{title}{Procedure for systematically tuning up cross-talk in
  the cross-resonance gate}.
\newblock \emph{\bibinfo{journal}{Phys. Rev. A}} \textbf{\bibinfo{volume}{93}},
  \bibinfo{pages}{060302} (\bibinfo{year}{2016}).
\newblock \urlprefix\url{https://link.aps.org/doi/10.1103/PhysRevA.93.060302}.

\bibitem{malekakhlagh2020firstprinciples}
\bibinfo{author}{Malekakhlagh, M.}, \bibinfo{author}{Magesan, E.} \&
  \bibinfo{author}{McKay, D.~C.}
\newblock \bibinfo{title}{First-principles analysis of cross-resonance gate
  operation}.
\newblock \emph{\bibinfo{journal}{Phys. Rev. A}}
  \textbf{\bibinfo{volume}{102}}, \bibinfo{pages}{042605}
  (\bibinfo{year}{2020}).
\newblock \urlprefix\url{https://link.aps.org/doi/10.1103/PhysRevA.102.042605}.

\bibitem{sundaresan2020reducing}
\bibinfo{author}{Sundaresan, N.} \emph{et~al.}
\newblock \bibinfo{title}{Reducing unitary and spectator errors in cross
  resonance with optimized rotary echoes}.
\newblock \emph{\bibinfo{journal}{PRX Quantum}} \textbf{\bibinfo{volume}{1}},
  \bibinfo{pages}{020318} (\bibinfo{year}{2020}).
\newblock
  \urlprefix\url{https://journals.aps.org/prxquantum/abstract/10.1103/PRXQuantum.1.020318}.

\bibitem{itoko2024three}
\bibinfo{author}{Itoko, T.}, \bibinfo{author}{Malekakhlagh, M.},
  \bibinfo{author}{Kanazawa, N.} \& \bibinfo{author}{Takita, M.}
\newblock \bibinfo{title}{Three-qubit parity gate via simultaneous
  cross-resonance drives}.
\newblock \emph{\bibinfo{journal}{Physical Review Applied}}
  \textbf{\bibinfo{volume}{21}}, \bibinfo{pages}{034018}
  (\bibinfo{year}{2024}).
\newblock
  \urlprefix\url{https://journals.aps.org/prapplied/abstract/10.1103/PhysRevApplied.21.034018}.

\bibitem{PhysRevA.105.032435}
\bibinfo{author}{Chen, S.}, \bibinfo{author}{Zhou, S.}, \bibinfo{author}{Seif,
  A.} \& \bibinfo{author}{Jiang, L.}
\newblock \bibinfo{title}{Quantum advantages for pauli channel estimation}.
\newblock \emph{\bibinfo{journal}{Phys. Rev. A}}
  \textbf{\bibinfo{volume}{105}}, \bibinfo{pages}{032435}
  (\bibinfo{year}{2022}).
\newblock \urlprefix\url{https://link.aps.org/doi/10.1103/PhysRevA.105.032435}.

\bibitem{knill2008randomized}
\bibinfo{author}{Knill, E.} \emph{et~al.}
\newblock \bibinfo{title}{Randomized benchmarking of quantum gates}.
\newblock \emph{\bibinfo{journal}{Phys. Rev. A}} \textbf{\bibinfo{volume}{77}},
  \bibinfo{pages}{012307} (\bibinfo{year}{2008}).
\newblock \urlprefix\url{https://link.aps.org/doi/10.1103/PhysRevA.77.012307}.

\bibitem{magesan2011scalable}
\bibinfo{author}{Magesan, E.}, \bibinfo{author}{Gambetta, J.~M.} \&
  \bibinfo{author}{Emerson, J.}
\newblock \bibinfo{title}{Scalable and robust randomized benchmarking of
  quantum processes}.
\newblock \emph{\bibinfo{journal}{Phys. Rev. Lett.}}
  \textbf{\bibinfo{volume}{106}}, \bibinfo{pages}{180504}
  (\bibinfo{year}{2011}).
\newblock
  \urlprefix\url{https://link.aps.org/doi/10.1103/PhysRevLett.106.180504}.

\bibitem{chen2023learnability}
\bibinfo{author}{Chen, S.} \emph{et~al.}
\newblock \bibinfo{title}{The learnability of pauli noise}.
\newblock \emph{\bibinfo{journal}{Nature Communications}}
  \textbf{\bibinfo{volume}{14}}, \bibinfo{pages}{52} (\bibinfo{year}{2023}).

\bibitem{seif2024suppressing}
\bibinfo{author}{Seif, A.} \emph{et~al.}
\newblock \bibinfo{title}{Suppressing correlated noise in quantum computers via
  context-aware compiling}.
\newblock In \emph{\bibinfo{booktitle}{Proceedings of the 51st Annual
  International Symposium on Computer Architecture}}
  (\bibinfo{organization}{IEEE}, \bibinfo{year}{2024}).

\bibitem{gambetta2012simulrb}
\bibinfo{author}{Gambetta, J.~M.} \emph{et~al.}
\newblock \bibinfo{title}{Characterization of addressability by simultaneous
  randomized benchmarking}.
\newblock \emph{\bibinfo{journal}{Phys. Rev. Lett.}}
  \textbf{\bibinfo{volume}{109}}, \bibinfo{pages}{240504}
  (\bibinfo{year}{2012}).
\newblock
  \urlprefix\url{https://link.aps.org/doi/10.1103/PhysRevLett.109.240504}.

\bibitem{shabani2011esitmation}
\bibinfo{author}{Shabani, A.}, \bibinfo{author}{Mohseni, M.},
  \bibinfo{author}{Lloyd, S.}, \bibinfo{author}{Kosut, R.~L.} \&
  \bibinfo{author}{Rabitz, H.}
\newblock \bibinfo{title}{Estimation of many-body quantum hamiltonians via
  compressive sensing}.
\newblock \emph{\bibinfo{journal}{Phys. Rev. A}} \textbf{\bibinfo{volume}{84}},
  \bibinfo{pages}{012107} (\bibinfo{year}{2011}).
\newblock \urlprefix\url{https://link.aps.org/doi/10.1103/PhysRevA.84.012107}.

\bibitem{dasilva2011practical}
\bibinfo{author}{da~Silva, M.~P.}, \bibinfo{author}{Landon-Cardinal, O.} \&
  \bibinfo{author}{Poulin, D.}
\newblock \bibinfo{title}{Practical characterization of quantum devices without
  tomography}.
\newblock \emph{\bibinfo{journal}{Phys. Rev. Lett.}}
  \textbf{\bibinfo{volume}{107}}, \bibinfo{pages}{210404}
  (\bibinfo{year}{2011}).
\newblock
  \urlprefix\url{https://link.aps.org/doi/10.1103/PhysRevLett.107.210404}.

\bibitem{Note1}
\bibinfo{note}{Since we are only learning symmetrized Pauli fidelities, one
  might not need to go through all $3^w$ Pauli basis measurements. However, a
  lower bound of $\Omega (2^w)$ measurements is still inevitable in this
  setting~\cite {chen2024tight}.}

\bibitem{ali2024native}
\bibinfo{author}{Wei, K.~X.} \emph{et~al.}
\newblock \bibinfo{title}{Native two-qubit gates in fixed-coupling,
  fixed-frequency transmons beyond cross-resonance interaction}.
\newblock \emph{\bibinfo{journal}{PRX Quantum}} \textbf{\bibinfo{volume}{5}},
  \bibinfo{pages}{020338} (\bibinfo{year}{2024}).
\newblock \urlprefix\url{https://link.aps.org/doi/10.1103/PRXQuantum.5.020338}.

\bibitem{heraldedstateprep}
\bibinfo{author}{Johnson, J.~E.} \emph{et~al.}
\newblock \bibinfo{title}{Heralded state preparation in a superconducting
  qubit}.
\newblock \emph{\bibinfo{journal}{Phys. Rev. Lett.}}
  \textbf{\bibinfo{volume}{109}}, \bibinfo{pages}{050506}
  (\bibinfo{year}{2012}).
\newblock
  \urlprefix\url{https://link.aps.org/doi/10.1103/PhysRevLett.109.050506}.

\bibitem{pino2021demonstration}
\bibinfo{author}{Pino, J.~M.} \emph{et~al.}
\newblock \bibinfo{title}{Demonstration of the trapped-ion quantum ccd computer
  architecture}.
\newblock \emph{\bibinfo{journal}{Nature}} \textbf{\bibinfo{volume}{592}},
  \bibinfo{pages}{209--213} (\bibinfo{year}{2021}).

\bibitem{bluvstein2024logical}
\bibinfo{author}{Bluvstein, D.} \emph{et~al.}
\newblock \bibinfo{title}{Logical quantum processor based on reconfigurable
  atom arrays}.
\newblock \emph{\bibinfo{journal}{Nature}} \textbf{\bibinfo{volume}{626}},
  \bibinfo{pages}{58--65} (\bibinfo{year}{2024}).

\bibitem{zhou2018achieving}
\bibinfo{author}{Zhou, S.}, \bibinfo{author}{Zhang, M.},
  \bibinfo{author}{Preskill, J.} \& \bibinfo{author}{Jiang, L.}
\newblock \bibinfo{title}{Achieving the heisenberg limit in quantum metrology
  using quantum error correction}.
\newblock \emph{\bibinfo{journal}{Nature communications}}
  \textbf{\bibinfo{volume}{9}}, \bibinfo{pages}{78} (\bibinfo{year}{2018}).

\bibitem{carignan2023error}
\bibinfo{author}{Carignan-Dugas, A.} \emph{et~al.}
\newblock \bibinfo{title}{The error reconstruction and compiled calibration of
  quantum computing cycles}.
\newblock \emph{\bibinfo{journal}{arXiv preprint arXiv:2303.17714}}
  (\bibinfo{year}{2023}).

\bibitem{harper2019statistical}
\bibinfo{author}{Harper, R.}, \bibinfo{author}{Hincks, I.},
  \bibinfo{author}{Ferrie, C.}, \bibinfo{author}{Flammia, S.~T.} \&
  \bibinfo{author}{Wallman, J.~J.}
\newblock \bibinfo{title}{Statistical analysis of randomized benchmarking}.
\newblock \emph{\bibinfo{journal}{Physical Review A}}
  \textbf{\bibinfo{volume}{99}}, \bibinfo{pages}{052350}
  (\bibinfo{year}{2019}).

\end{thebibliography}


\begin{thebibliography}{10}
\expandafter\ifx\csname url\endcsname\relax
  \def\url#1{\texttt{#1}}\fi
\expandafter\ifx\csname urlprefix\endcsname\relax\def\urlprefix{URL }\fi
\providecommand{\bibinfo}[2]{#2}
\providecommand{\eprint}[2][]{\url{#2}}

\bibitem{chen2022quantum}
\bibinfo{author}{Chen, S.}, \bibinfo{author}{Zhou, S.}, \bibinfo{author}{Seif,
  A.} \& \bibinfo{author}{Jiang, L.}
\newblock \bibinfo{title}{Quantum advantages for pauli channel estimation}.
\newblock \emph{\bibinfo{journal}{Physical Review A}}
  \textbf{\bibinfo{volume}{105}}, \bibinfo{pages}{032435}
  (\bibinfo{year}{2022}).

\bibitem{chen2024tight}
\bibinfo{author}{Chen, S.}, \bibinfo{author}{Oh, C.}, \bibinfo{author}{Zhou,
  S.}, \bibinfo{author}{Huang, H.-Y.} \& \bibinfo{author}{Jiang, L.}
\newblock \bibinfo{title}{Tight bounds on pauli channel learning without
  entanglement}.
\newblock \emph{\bibinfo{journal}{Physical Review Letters}}
  \textbf{\bibinfo{volume}{132}}, \bibinfo{pages}{180805}
  (\bibinfo{year}{2024}).

\bibitem{chen2023futility}
\bibinfo{author}{Chen, S.} \& \bibinfo{author}{Gong, W.}
\newblock \bibinfo{title}{Futility and utility of a few ancillas for pauli
  channel learning}.
\newblock \emph{\bibinfo{journal}{arXiv preprint arXiv:2309.14326}}
  (\bibinfo{year}{2023}).

\bibitem{bennett1992communication}
\bibinfo{author}{Bennett, C.~H.} \& \bibinfo{author}{Wiesner, S.~J.}
\newblock \bibinfo{title}{Communication via one-and two-particle operators on
  einstein-podolsky-rosen states}.
\newblock \emph{\bibinfo{journal}{Physical review letters}}
  \textbf{\bibinfo{volume}{69}}, \bibinfo{pages}{2881} (\bibinfo{year}{1992}).

\bibitem{hoeffding1994probability}
\bibinfo{author}{Hoeffding, W.}
\newblock \bibinfo{title}{Probability inequalities for sums of bounded random
  variables}.
\newblock \emph{\bibinfo{journal}{The collected works of Wassily Hoeffding}}
  \bibinfo{pages}{409--426} (\bibinfo{year}{1994}).

\bibitem{wallman2016noise}
\bibinfo{author}{Wallman, J.~J.} \& \bibinfo{author}{Emerson, J.}
\newblock \bibinfo{title}{Noise tailoring for scalable quantum computation via
  randomized compiling}.
\newblock \emph{\bibinfo{journal}{Physical Review A}}
  \textbf{\bibinfo{volume}{94}}, \bibinfo{pages}{052325}
  (\bibinfo{year}{2016}).

\bibitem{hashim2020randomized}
\bibinfo{author}{Hashim, A.} \emph{et~al.}
\newblock \bibinfo{title}{Randomized compiling for scalable quantum computing
  on a noisy superconducting quantum processor}.
\newblock \emph{\bibinfo{journal}{arXiv preprint arXiv:2010.00215}}
  (\bibinfo{year}{2020}).

\bibitem{van2023probabilistic}
\bibinfo{author}{van Den~Berg, E.}, \bibinfo{author}{Minev, Z.~K.},
  \bibinfo{author}{Kandala, A.} \& \bibinfo{author}{Temme, K.}
\newblock \bibinfo{title}{Probabilistic error cancellation with sparse
  pauli--lindblad models on noisy quantum processors}.
\newblock \emph{\bibinfo{journal}{Nature Physics}} \bibinfo{pages}{1--6}
  (\bibinfo{year}{2023}).

\bibitem{chen2023learnability}
\bibinfo{author}{Chen, S.} \emph{et~al.}
\newblock \bibinfo{title}{The learnability of pauli noise}.
\newblock \emph{\bibinfo{journal}{Nature Communications}}
  \textbf{\bibinfo{volume}{14}}, \bibinfo{pages}{52} (\bibinfo{year}{2023}).

\bibitem{flammia2020efficient}
\bibinfo{author}{Flammia, S.~T.} \& \bibinfo{author}{Wallman, J.~J.}
\newblock \bibinfo{title}{Efficient estimation of pauli channels}.
\newblock \emph{\bibinfo{journal}{ACM Transactions on Quantum Computing}}
  \textbf{\bibinfo{volume}{1}}, \bibinfo{pages}{1--32} (\bibinfo{year}{2020}).

\bibitem{huang2022foundations}
\bibinfo{author}{Huang, H.-Y.}, \bibinfo{author}{Flammia, S.~T.} \&
  \bibinfo{author}{Preskill, J.}
\newblock \bibinfo{title}{Foundations for learning from noisy quantum
  experiments}.
\newblock \emph{\bibinfo{journal}{arXiv preprint arXiv:2204.13691}}
  (\bibinfo{year}{2022}).

\bibitem{erhard2019characterizing}
\bibinfo{author}{Erhard, A.} \emph{et~al.}
\newblock \bibinfo{title}{Characterizing large-scale quantum computers via
  cycle benchmarking}.
\newblock \emph{\bibinfo{journal}{Nature communications}}
  \textbf{\bibinfo{volume}{10}}, \bibinfo{pages}{5347} (\bibinfo{year}{2019}).

\bibitem{berg2023techniques}
\bibinfo{author}{van~den Berg, E.} \& \bibinfo{author}{Wocjan, P.}
\newblock \bibinfo{title}{Techniques for learning sparse pauli-lindblad noise
  models}.
\newblock \emph{\bibinfo{journal}{arXiv preprint arXiv:2311.15408}}
  (\bibinfo{year}{2023}).

\bibitem{gambetta2012simulrb}
\bibinfo{author}{Gambetta, J.~M.} \emph{et~al.}
\newblock \bibinfo{title}{Characterization of addressability by simultaneous
  randomized benchmarking}.
\newblock \emph{\bibinfo{journal}{Phys. Rev. Lett.}}
  \textbf{\bibinfo{volume}{109}}, \bibinfo{pages}{240504}
  (\bibinfo{year}{2012}).
\newblock
  \urlprefix\url{https://link.aps.org/doi/10.1103/PhysRevLett.109.240504}.

\bibitem{harper2019statistical}
\bibinfo{author}{Harper, R.}, \bibinfo{author}{Hincks, I.},
  \bibinfo{author}{Ferrie, C.}, \bibinfo{author}{Flammia, S.~T.} \&
  \bibinfo{author}{Wallman, J.~J.}
\newblock \bibinfo{title}{Statistical analysis of randomized benchmarking}.
\newblock \emph{\bibinfo{journal}{Physical Review A}}
  \textbf{\bibinfo{volume}{99}}, \bibinfo{pages}{052350}
  (\bibinfo{year}{2019}).

\bibitem{2020SciPy-NMeth}
\bibinfo{author}{Virtanen, P.} \emph{et~al.}
\newblock \bibinfo{title}{{{SciPy} 1.0: Fundamental Algorithms for Scientific
  Computing in Python}}.
\newblock \emph{\bibinfo{journal}{Nature Methods}}
  \textbf{\bibinfo{volume}{17}}, \bibinfo{pages}{261--272}
  (\bibinfo{year}{2020}).

\bibitem{seif2024suppressing}
\bibinfo{author}{Seif, A.} \emph{et~al.}
\newblock \bibinfo{title}{Suppressing correlated noise in quantum computers via
  context-aware compiling}.
\newblock In \emph{\bibinfo{booktitle}{Proceedings of the 51st Annual
  International Symposium on Computer Architecture}}
  (\bibinfo{organization}{IEEE}, \bibinfo{year}{2024}).

\bibitem{Tripathi_Operation_2019}
\bibinfo{author}{Tripathi, V.}, \bibinfo{author}{Khezri, M.} \&
  \bibinfo{author}{Korotkov, A.~N.}
\newblock \bibinfo{title}{Operation and intrinsic error budget of a two-qubit
  cross-resonance gate}.
\newblock \emph{\bibinfo{journal}{Phys. Rev. A}}
  \textbf{\bibinfo{volume}{100}}, \bibinfo{pages}{012301}
  (\bibinfo{year}{2019}).
\newblock \urlprefix\url{https://link.aps.org/doi/10.1103/PhysRevA.100.012301}.

\bibitem{Magesan_Effective_2020}
\bibinfo{author}{Magesan, E.} \& \bibinfo{author}{Gambetta, J.~M.}
\newblock \bibinfo{title}{Effective hamiltonian models of the cross-resonance
  gate}.
\newblock \emph{\bibinfo{journal}{Phys. Rev. A}}
  \textbf{\bibinfo{volume}{101}}, \bibinfo{pages}{052308}
  (\bibinfo{year}{2020}).
\newblock \urlprefix\url{https://link.aps.org/doi/10.1103/PhysRevA.101.052308}.

\bibitem{Malekakhlagh_First_2020}
\bibinfo{author}{Malekakhlagh, M.}, \bibinfo{author}{Magesan, E.} \&
  \bibinfo{author}{McKay, D.~C.}
\newblock \bibinfo{title}{First-principles analysis of cross-resonance gate
  operation}.
\newblock \emph{\bibinfo{journal}{Phys. Rev. A}}
  \textbf{\bibinfo{volume}{102}}, \bibinfo{pages}{042605}
  (\bibinfo{year}{2020}).
\newblock \urlprefix\url{https://link.aps.org/doi/10.1103/PhysRevA.102.042605}.

\bibitem{leung2002simulation}
\bibinfo{author}{Leung, D.}
\newblock \bibinfo{title}{Simulation and reversal of n-qubit hamiltonians using
  hadamard matrices}.
\newblock \emph{\bibinfo{journal}{Journal of Modern Optics}}
  \textbf{\bibinfo{volume}{49}}, \bibinfo{pages}{1199--1217}
  (\bibinfo{year}{2002}).

\bibitem{rotteler2006equivalence}
\bibinfo{author}{Rotteler, M.} \& \bibinfo{author}{Wocjan, P.}
\newblock \bibinfo{title}{Equivalence of decoupling schemes and orthogonal
  arrays}.
\newblock \emph{\bibinfo{journal}{IEEE transactions on information theory}}
  \textbf{\bibinfo{volume}{52}}, \bibinfo{pages}{4171--4181}
  (\bibinfo{year}{2006}).

\bibitem{zhou2022quantum}
\bibinfo{author}{Zhou, Z.}, \bibinfo{author}{Sitler, R.}, \bibinfo{author}{Oda,
  Y.}, \bibinfo{author}{Schultz, K.} \& \bibinfo{author}{Quiroz, G.}
\newblock \bibinfo{title}{Quantum crosstalk robust quantum control}.
\newblock \emph{\bibinfo{journal}{Phys. Rev. Lett.}}
  \textbf{\bibinfo{volume}{131}}, \bibinfo{pages}{210802}
  (\bibinfo{year}{2023}).
\newblock
  \urlprefix\url{https://link.aps.org/doi/10.1103/PhysRevLett.131.210802}.

\bibitem{Shirizly2024dissipative}
\bibinfo{author}{Shirizly, L.}, \bibinfo{author}{Misguich, G.} \&
  \bibinfo{author}{Landa, H.}
\newblock \bibinfo{title}{Dissipative dynamics of graph-state stabilizers with
  superconducting qubits}.
\newblock \emph{\bibinfo{journal}{Phys. Rev. Lett.}}
  \textbf{\bibinfo{volume}{132}}, \bibinfo{pages}{010601}
  (\bibinfo{year}{2024}).
\newblock
  \urlprefix\url{https://link.aps.org/doi/10.1103/PhysRevLett.132.010601}.

\end{thebibliography}

\end{document}


\title{Entanglement enhanced learning of quantum processes at scale: Supplemental Material}
\author{Alireza Seif}
\thanks{These authors contributed equally to this work.}
\affiliation{IBM Quantum, IBM T.J. Watson Research Center, Yorktown Heights, NY, USA}
\author{Senrui Chen}
\thanks{These authors contributed equally to this work.}
\affiliation{Pritzker School of Molecular Engineering, University of Chicago, Chicago, IL, USA}
\author{Swarnadeep Majumder}
\affiliation{IBM Quantum, IBM T.J. Watson Research Center, Yorktown Heights, NY, USA}
\author{Haoran Liao}
\affiliation{%
Department of Physics, University of California, Berkeley, CA 94720, USA
}%
\author{Derek S Wang}
\affiliation{IBM Quantum, IBM T.J. Watson Research Center, Yorktown Heights, NY, USA}
\author{Moein Malekakhlagh}
\affiliation{IBM Quantum, IBM T.J. Watson Research Center, Yorktown Heights, NY, USA}
\author{Ali Javadi-Abhari}
\affiliation{IBM Quantum, IBM T.J. Watson Research Center, Yorktown Heights, NY, USA}
\author{Liang Jiang}
\affiliation{Pritzker School of Molecular Engineering, University of Chicago, Chicago, IL, USA}
\author{Zlatko K Minev}
\affiliation{IBM Quantum, IBM T.J. Watson Research Center, Yorktown Heights, NY, USA}
\maketitle
\tableofcontents

\section{Pauli-transfer-matrix (PTM) representation}

We have defined Pauli group, Pauli channels, and Bell states in Methods Sec.~A. 
Here we introduce the \emph{Pauli-transfer-matrix} (PTM) representation to further simplify notations.
Our notations follows, e.g., Ref.~\cite{chen2022quantum}.
A linear operator $O$ acting on a $2^n$-dimensional Hilbert space can be viewed as a vector in a $4^n$-dimensional Hilbert space. We denote this vectorization of $O$ as $\lket{O}$ and the corresponding Hermitian conjugate as $\lbra{O}$. The inner product within this space is the \emph{Hilbert-Schmidt} product defined as $\lbraket{A}{B}\coleq \Tr(A^\dagger B)$. The \emph{normalized Pauli operators} $\{\sigma_a\coleq P_a/\sqrt{2^{n}},~a\in\mbb Z_2^{2n}\}$ forms an orthonormal basis for this space. 
In the PTM representation, a superoperator (\textit{i.e.}, quantum channel) becomes an operator acting on the $4^n$-dimensional Hilbert space, sometimes called the Pauli transfer operator. 
Explicitly, we have $\lket{\Lambda(\rho)} = \Lambda^\ptm\lket{\rho} \equiv \Lambda\lket{\rho}$,
where we use the same notation to denote a channel and its Pauli transfer operator, which should be clear from the context.
Specifically, a general Pauli channel $\Lambda$ has the following Pauli transfer operator
$$
\Lambda = \sum_{a\in\mbb Z_2^{2n}} \lambda_a\lketbra{\sigma_a}{\sigma_a}.
$$
In the PTM representation, the $n$-qubit Bell states are given by
\begin{equation}
    \lket{{\Phi}_v} = \mathcal P_v \otimes\mathds 1\lket{{\Phi}_+} = \frac{1}{4^n}\sum_{u\in\mbb Z_2^{2n}}(-1)^{\expval{u,v}}\lket{P_u\otimes P_u^T},\quad\forall v\in \mbb Z_2^{2n},
\end{equation}
which forms a complete basis on $\mc H_{2^{2n}}$. 
Also, we use $\lket{j}$ to denote the vectorization of the computational basis state $\ketbra{j}{j}$ for $j=1\cdots2^n-1$.

\section{Pauli channel learning in black-box model}

In this section, we review the Pauli channel learning task in the black-box model. That is, a Pauli channel $\Lambda$ given as a black box for one to query. This is to be distinguished with the gate-dependent Pauli noise model that we will consider in the next section. 
The task we study is the following

\begin{task}[($\varepsilon,\delta$)-Pauli eigenvalue estimation]\label{task1}
    Given parameters $\varepsilon,\delta\in(0,1)$, after one makes $N$ black-box queries to an $n$-qubit Pauli channel $\Lambda$, a question $a\in\Pn$ is given and one is asked to generate an estimator $\hat\lambda_a$ to $\lambda_a$ such that $|\lambda_a-\hat\lambda_a|\le\varepsilon$ with at least $1-\delta$ probability for any question $a$. $N$ is called the sample complexity of the protocol.
\end{task}

Task~\ref{task1} requires estimating only one but arbitrary Pauli eigenvalue $\lambda_a$. Note that, all quantum learning procedures must complete before the choice of $a$ is revealed, after which only classical post-processing is allowed.
Importantly, this task is proved to be hard for entanglement-free learning (\sysb{}) protocols~\cite{chen2024tight} (see also Ref.~\cite{chen2023futility} for a slightly more restricted class of protocols).
For a constantly high success probability, this sample complexity scales as $\Omega(2^n/\varepsilon^2)$, which quickly becomes infeasible with increasing system size. In contrast, entanglement-enhanced learning (\eab{}) with access to perfect (noisy) quantum memory can complete this task with constant (weakly exponential) sample complexity, respectively. In the following, we first discuss \eab{} protocols with perfect Bell pairs, then introduce SPAM-robust \eab{} with noisy Bell pairs, and finally review the hardness results for \sysb{} protocols. These are the theoretic foundations for our hypothesis testing experiment presented in the main text Sec. III.

\subsection{Entanglement-enhanced learning with perfect Bell pairs}
Let $\ket{\phi_+} = (\ket{00}+\ket{11})/\sqrt2$ be one canonical Bell pair. If one applies Pauli $\{I,X,Y,Z\}$ on the first qubit and identity on the second, the output states will be the four Bell states $\{\ket{\phi_+},\ket{\psi_+},i\ket{\psi_-},\ket{\phi_-}\}$ where
\begin{equation}
    \ket{\phi_{\pm}} = \frac{1}{\sqrt2}(\ket{00}\pm\ket{11}),\quad \ket{\psi_{\pm}} = \frac{1}{\sqrt2}(\ket{01}\pm\ket{10}).
\end{equation}
Since the four Bell states form an orthogonal basis for the $2$-qubit Hilbert space, a projective measurement onto that basis uniquely determines the Pauli being applied on the first qubit. This procedure is known as quantum superdense coding~\cite{bennett1992communication}. Generalizing this to a $2n$-qubit system yields a Pauli channel learning scheme as follows~\cite{chen2022quantum}:

Let $S$, $A$ be two $n$-qubit quantum systems. Define $\ket{\Phi_+}_{SA}\coleq\ket{\phi_+}_{S_1A_1}\otimes\cdots\otimes\ket{\phi_+}_{S_nA_n}$ as the canonical Bell state. It is not hard to see the density matrix of $\ketbra{\Phi_+}{\Phi_+}$ can be expressed as
\begin{equation}
    \Phi_+ \coleq \ketbra{\Phi_+}{\Phi_+}= \frac{1}{4^n}\sum_{a\in\Pn}P_a\otimes P_a^\T.
\end{equation}
The $2n$-qubit Bell basis is denoted by $\{\ket{\Phi_b}\}_{b\in\Pn}$ where $\ket{\Phi_b}\coleq P_b\otimes I\ket{\Phi_+}$. The orthogonality between different $\ket{\Phi_b}$ can be readily checked. Alternatively,
\begin{equation}
    \Phi_b \coleq \ketbra{\Phi_b}{\Phi_b} = \frac{1}{4^n}\sum_{a\in\Pn}(-1)^{\expval{a,b}}P_a\otimes P_a^\T.
\end{equation}
Consider the following experiment: Prepare $\ket{\Phi_+}_{SA}$, apply $\Lambda$ on system $S$, then measure the whole system with the Bell basis. The probability of getting measurement outcome $b$ is
\begin{equation}\label{eq:deri_bell_samp}
    \Pr[b] = \Tr[\Phi_b (\Lambda_S\otimes\mathds 1_A(\Phi_+))] = \frac{1}{4^n}\Tr\left[\Phi_b\left(\sum_{a\in\Pn}\lambda_aP_a\otimes P_a^\T\right)\right]
    = \frac{1}{4^n}\sum_{a\in\Pn}(-1)^{\expval{a,b}}\lambda_a.
\end{equation}
Comparing this to the Walsh-Hadamard transform, one can realize that $\Pr[b]$ is exactly the Pauli error rates $p$ of $\Lambda$. Specifically, we have
\begin{equation}\label{eq:unbiased}
    \lambda_a = \sum_{b\in\Pn}(-1)^{\expval{a,b}}\Pr[b].
\end{equation}
Now we can describe the learning protocol as follows: Repeat the above experiment $N$ times and collect the measurement outcomes $\bm b=\{b^{(i)}\}_{i=1}^N$. Then, for any $\lambda_a$, one computes the following estimator
\begin{equation}
\hat\lambda_a[
\bm b
]\coleq \frac1N \sum_{i=1}^N\hat\lambda_a[b^{(i)}],\quad\text{where}~\hat\lambda_a[b]\coleq(-1)^{\langle{a,b}\rangle}.
\end{equation}
This estimator is unbiased as Eq.~\eqref{eq:unbiased} says $\mbb E_{b\sim p}(-1)^{\expval{a,b}} = \lambda_a$. Furthermore, the single-shot estimator is bounded between $\pm 1$, so Hoeffding's inequality~\cite{hoeffding1994probability} immediately gives that
\begin{equation}
    \Pr[\left|\hat{\lambda}_a[\bm b]-\lambda_a\right|\ge \varepsilon]\le 2\exp(-\frac12 N\varepsilon^2).
\end{equation}
Therefore, this protocol estimates any $\lambda_a$ to $\varepsilon$ precision with success probability at least $1-\delta$ using the following number of samples (i.e., queries to $\Lambda$)
\begin{equation}
    N = 2\varepsilon^{-2}\log(2\delta^{-1}).
\end{equation}
Crucially, this sample complexity is independent of the system size.

\subsection{Entanglement-enhanced learning with noisy Bell pairs}

In current noisy quantum computing platforms, we do not have access to perfect Bell pairs and measurement. However, with a simple calibration step, the above learning protocol can be made robust against state preparation and measurement (SPAM) error.
Suppose the noisy realization of the Bell state $\Phi_+$ and Bell measurement POVM $\{\Phi_a\}_a$ can be written as
\begin{equation}
    \widetilde{\Phi}_+ = \mc E^\s(\Phi_+);\quad \widetilde{\Phi}_a = (\mc E^\m)^\dagger(\Phi_a),
\end{equation}
for some CPTP maps $\mc E^\s,~\mc E^\m$.
We assume the ability to perform noiseless Pauli gates (or, less stringently, all Pauli gates have a gate-independent noise channel). This assumption is relevant for many existing experimental platforms and is standard for Pauli twirling-based methods~\cite{wallman2016noise,hashim2020randomized,van2023probabilistic}. With this assumption, one can always twirl the noise channel into a Pauli channel. In the following, we assume the Pauli twirling has been done and both $\mc E^\s$, $\mc E^\m$ are Pauli channels.
Let $\mc E\coleq \mc E^\m\circ\mc E^\s$ be their composition and denote the Pauli eigenvalue of $\mc E$ as $\xi_a$.

\medskip

Consider the following two experiments:
\begin{enumerate}
    \item Prepare $\widetilde{\Phi}^{SA}_+$. Then measure with $\{\widetilde{\Phi}^{SA}_a\}$.
    \item Prepare $\widetilde{\Phi}^{SA}_+$. Apply $\Lambda$ on system $S$. Then measure with $\{\widetilde{\Phi}^{SA}_a\}$.
\end{enumerate}
The distribution of measurement outcome of experiment 1 can be calculated as
\begin{equation}
\begin{aligned}
    \Pr[b]_1 &= \Tr[\Phi_b (\mc E^\m\circ\mc E^\s)(\Phi_+)]\\ 
    &=\frac{1}{4^n}\Tr[\Phi_b \left(\mc E\left(\sum_{a\in\Pn}P_a\otimes P_a^\T\right)\right)]\\
    &=\frac{1}{4^n}\Tr[\Phi_b \left(\sum_{a\in\Pn}\xi_{a,a}P_a\otimes P_a^\T\right)]\\
    &= \frac{1}{4^n}\sum_{a\in\Pn}(-1)^{\expval{a,b}}\xi_{a,a}.
\end{aligned}
\end{equation}
where $\xi_{a,a}$ is the Pauli eigenvalue of $\mc E$ corresponding to $P_a\otimes P_a$. Similarly, experiment 2 yields 
\begin{equation}
\begin{aligned}
    \Pr[b]_2 &= \Tr[\Phi_b (\mc E^\m\circ(\Lambda\otimes\mathds 1)\circ\mc E^\s)(\Phi_+)]\\
    &=\frac{1}{4^n}\sum_{a\in\Pn}(-1)^{\expval{a,b}}\xi_{a,a}\lambda_a,  
\end{aligned}
\end{equation}
where we use the fact that all Pauli channels commute.
Compare these to Eq.~\eqref{eq:deri_bell_samp}, we see that one can use similar estimation procedure as the noiseless case to efficiently estimate $\xi_{a,a}$ and $\xi_{a,a}\lambda_a$ to small additive error with high probability from experiment 1 and 2, respectively. By dividing the latter estimator with the former, we can obtain an estimator for $\lambda_a$ that is independent of $\xi_{a,a}$ in the asymptotic limit. 

More formally, consider the following protocol: Repeat $N_1$ times experiment $1$ and collect measurement outcomes $\bm {b}=\{b^{(i)}\}_{i=1}^{N_1}$; Then repeat $N_2$ times experiment $2$ and collect measurement outcomes $\bm {c}=\{c^{(i)}\}_{i=1}^{N_2}$; Now, for any $\lambda_a$, compute the following estimators
\begin{equation}
    \hat{\lambda}^{(1)}_a \coleq \frac1{N_1}\sum_{i=1}^{N_1}(-1)^{\langle{a,b^{(i)}}\rangle},\quad
    \hat{\lambda}^{(2)}_a \coleq \frac1{N_2}\sum_{i=1}^{N_2}(-1)^{\langle{a,c^{(i)}}\rangle},\quad
    \hat{\lambda}_a \coleq \frac{\hat{\lambda}^{(2)}_a}{\hat{\lambda}^{(1)}_a}.
\end{equation}
It is easy to see that $\hat{\lambda}_a$ is a consistent (\textit{i.e.}, asymptotically unbiased) estimator for $\lambda_a$.
Let us define
\begin{equation}
    \hat{\varepsilon}_a^{(1)}\coleq \hat{\lambda}_a^{(1)}-\xi_{a,a},\quad \hat{\varepsilon}_a^{(2)}\coleq \hat{\lambda}_a^{(2)} - \xi_{a,a}\lambda_a.
\end{equation}
Thanks to the Hoeffding's inequality, if we choose $$N_1 = 2\varepsilon_1^{-2}\log(2\delta_1^{-1}),\quad N_2 = 2\varepsilon_2^{-2}\log(2\delta_2^{-1}),$$
we have $|\hat{\varepsilon}_a^{(1)}|\le\varepsilon_1$ with probability at least $1-\delta_1$ and $|\hat{\varepsilon}_a^{(2)}|\le\varepsilon_2$ with probability at least $1-\delta_2$. 
Thus, the following holds with probability at least $1-\delta_1-\delta_2$,
\begin{equation}
    \begin{aligned}
        |\hat{\lambda}_a - \lambda_a|&= \left|\frac{\xi_{a,a}\lambda_a+\hat{\varepsilon}_a^{(2)}}{\xi_{a,a}+\hat{\varepsilon}_a^{(1)}}-\lambda_a\right|
        = \left|\frac{\hat{\varepsilon}_a^{(2)}-\lambda_a\hat{\varepsilon}_a^{(1)}}{\xi_{a,a}+\hat{\varepsilon}_a^{(1)}}\right|
        \le \frac{\varepsilon_1+\varepsilon_2}{\xi_{a,a}-\varepsilon_1},
    \end{aligned}
\end{equation}
given that $\varepsilon_1<\xi_{a,a}$. 
Now, assuming that $\xi_{a,a}\ge F$ for all $a$, let us substitute
\begin{equation}
    \varepsilon_1=\varepsilon_2 = \frac13\varepsilon F,\quad \delta_1=\delta_2=\frac12\delta.
\end{equation}
Then, with probability no less than $1-\delta$, one has $|\hat{\lambda}_a-\lambda_a|\le \varepsilon$, accomplishing Task~\ref{task1}. The sample complexity is
\begin{equation}
    N_1 = N_2 = 18\varepsilon^{-2}F^{-2}\log(4\delta^{-1}).
\end{equation}
In practice, we expect the Bell state Pauli fidelity $\xi_{a,a}$ decays exponentially with the Pauli weight of $a$. Assume $F=f^n$ for some constant $f$ characterizing the per qubit fidelity, the sample complexity becomes 
\begin{equation}
N_1 = N_2 = O(f^{-2n}\varepsilon^{-2}\log\delta^{-1}),
\end{equation}
which depends exponentially with the number of qubits $n$. However, as long as $f$ is sufficiently close to $1$, this sample complexity will grow mildly and is manageable for a reasonable system size. Specifically, it will have a huge separation with the sample complexity lower bound from entanglement-free schemes. We see this separation in our experiment.

\subsection{Hardness for entanglement-free learning from hypothesis testing}

In this section, we review the hardness results for learning Pauli channels with \sysb{} protocols~\cite{chen2024tight}. The proof is based on reducing learning to a hypothesis testing task. In fact, this is exactly the first task conducted in our experiment as presented in the main text Sec. III. In below, we review the hypothesis testing task and its hardness for \sysb{} protocols.

\medskip

Given $\varepsilon\le1/6$, define the following set of Pauli channels
\begin{equation}
\begin{aligned}
    \Lambda_{a,\pm} &= \lketbra{\sigma_{\mb 0}}{\sigma_{\mb 0}} \pm 2\varepsilon\lketbra{\sigma_{a}}{\sigma_{a}},\quad\forall a\in{\sf P}^n~s.t. ~a\neq 0.\\
    \Lambda_{0} &=\lketbra{\sigma_{\mb 0}}{\sigma_{\mb 0}}.
\end{aligned}
\end{equation}
In terms of Pauli error rates, it is easy to verify that 
\begin{equation}
p_{a,\pm} = \frac{1\pm\varepsilon(-1)^{\expval{a,b}}}{4^n},\quad p_0 = \frac1{4^n}.   
\end{equation}
Therefore, $\Lambda_0$ is the depolarizing channel, while each $\Lambda_{a,\pm}$ can be viewed as a perturbed depolarizing channel.

\medskip

Now, consider the following game: A referee randomly sample $a\in\Pn\backslash\{I_n\}$ and $s\in\{\pm 1\}$ from a uniform distribution. Then, he chooses one of the following two Pauli channels with equal probability: (1) $\Lambda = \Lambda_0$ or (2) $\Lambda = \Lambda_{a,s}$. A player is then allowed to query $\Lambda$ using at most $N$ shots of measurement with a specific class of learning schemes and collect classical data. After that, the referee reveal the value of $a$. The player is then asked to guess whether (1) $\Lambda = \Lambda_0$ or (2) $\Lambda=\Lambda_{a,s}$. He succeeds if guessing correctly.
Note that, the player must finish all measurement before the revelation and can only do classical postprocessing afterwards. 
\begin{theorem}[Hardness for hypothesis testing with \sysb{}~\cite{chen2024tight}]\label{th:EF_hard}
    If the player uses an entanglement-free learning scheme, the average success probability is upper bounded by
    \begin{equation}
    \Pr[\mr{Success}]\le\frac12 + 30 N\varepsilon^2\frac{2^n}{4^n-1} = \frac12 + O(N\varepsilon^2/2^n).
    \end{equation}
    Here, the average is over the choices of $a$, $s$, and the two hypotheses of $\Lambda$.
\end{theorem}
\begin{proof}
    Given $a,s$ and an entanglement-free learning scheme, use $p_0(\mb o)$, $p_{a,s}(\mb o)$ to denote the probability distribution over measurement outcomes under hypothesis $(1)$, $(2)$, respectively.
    Now, for a fixed $a$, thanks to the revelation of $a$'s value, the player knows the measurement distribution is either $p_0$ or $\E_{s=\pm1}p_{a,s}$. 
    The success probability to distinguish between these two hypotheses is upper bounded by
    \begin{equation}
        \Pr[\text{Success}|a]\le\frac12(1+\mr{TVD}(p_0,\E_s p_{a,s})).
    \end{equation}
    Thus, the average success probability of distinguishing the two hypotheses is given by
    \begin{equation}
        \Pr[\text{Success}] = \E_a \Pr[\text{Success}|a] \le \frac12(1+\E_a\mr{TVD}(p_0,\E_s p_{a,s})).
    \end{equation}
    On the other hand, it is shown in the proof of \cite[Theorem 2]{chen2024tight} that
    \begin{equation}
        \E_a\mr{TVD}(p_0,\E_s p_{a,s})\le 4N\varepsilon^2\frac{2^n}{4^n-1}\left(1+2\sqrt{f(2\varepsilon)}\right),
    \end{equation}
    for some function $f$ such that $f(2\varepsilon)\le 44$ for $\varepsilon\in(0,1/6]$. Inserting this number completes the proof.
\end{proof}

By inserting $\varepsilon=1/6$ into the above theorem and requiring $n\ge 4$, one can obtain that
\begin{equation}
    \Pr[\mr{Success}] \le \frac12 + 0.43\times N 2^{-n}.
\end{equation}
which gives the \sysb{} success probability upper bound Eq.~(7) in the main text.

\medskip

Finally, since a learning protocol that solves Task~\ref{task1} can also solve the above hypothesis testing task with success probability at least $1-\delta$ (by simply estimating $\lambda_a)$, we immediately have the following corollary:
\begin{corollary}\cite{chen2024tight}\label{th:ef_task1}
    Any \sysb{} protocols need the following number of samples to complete Task~\ref{task1},
    \begin{equation}
        N \ge \frac{1}{60}(1-2\delta)(2^n-2^{-n})\varepsilon^{-2}
    \end{equation}
    given that $\varepsilon\in(0,1/6]$, $\delta\in(0,1/2)$.
\end{corollary}

\section{Proof for the correctness of \emeab{}}

In this section, we prove the \emeab{} protocol, described in Methods Sec B, gives a consistent (i.e., asymptotically unbiased) estimator for each Pauli fidelity $\lambda_a^{\mc G}$. Recall that, we make the following assumptions:
\begin{enumerate}
    \item Layers of single-qubit Clifford gates are noiseless. 
    \item The noises acting on the system and ancilla is time-stationary, Markovian, and have no crosstalk or contextual dependence between each other. Specifically, applying a multi-qubit Clifford gate $\mc G$ on the system and idling the ancilla is realized as
    $$
    \mc I^A\otimes\mc G^S \mapsto \widetilde{\mc I}^A\otimes\widetilde{\mc G}^S \coleq \Lambda_{\mr{anc}}^A\otimes{(\Lambda_{\mc G}\circ\mc G)}^S
    $$
    \item The initial state and measurement suffer from noise channel acting jointly on the system and ancilla, 
    $$\rho_0^{AS}\mapsto\mc E^\s(\rho_0^{AS}),\quad \mc M^{AS}\mapsto (\mc M\circ \mc E^\m)^{AS}.$$
\end{enumerate}

The first assumption are standard~\cite{wallman2016noise,hashim2020randomized,chen2023learnability,flammia2020efficient}.
The second assumption requires the system and ancilla to be well isolated, which can be ensured via dynamical decoupling, turning off certain couplers on superconducting platforms, or moving qubits far apart on ion or atom platforms, etc. The goal is to characterize the gate noise $\Lambda_{\mc G}$ in a SPAM-robust manner. Since we are going to apply Pauli twirling, we will only be interested in learning the Pauli eigenvalue of the Pauli twirl of $\Lambda_{\mc G}$, defined as $\{\lambda_a^{\mc G}\}$. We also denote the Pauli eigenvalue of $\lambda_{\mr{anc}}$ as $\{\lambda_a^\anc\}$.

As mentioned in the main text, it has been proved that not every $\lambda_a^{\mc G}$ can be learned SPAM-independently for a generic Clifford gate $\mc G$~\cite{chen2023learnability,huang2022foundations}, which is a fundamental limitation posted by model identifiability that is generic to any learning scheme.
Take $\mc G = \mr{CNOT}$ as an example, one can show that $\lambda_{XX},\lambda_{XI}$ cannot be individually learned SPAM-independently, while their product $\lambda_{XX}\lambda_{XI}$ can be. 
In our work, we only focus on learning a subset of learnable degrees of freedom of $\Lambda_{\mc G}$. 
Formally, we aim at learning the following quantities,
\begin{definition}[Symmetrized Pauli eigenvalues]
    Given an $n$-qubit Clifford gate $\mc G$, let $d_0$ be the smallest positive integer such that $\mc G^{d_0} = \mc I$. We define the \emph{symmetrized Pauli eigenvalues} as follows:
    \begin{equation}
        \bar{\lambda}_a\coleq \left(\prod_{k=0}^{d_0-1} \lambda_{\mc G^k(a)} \right)^{1/d_0},\quad \forall a\in\Pn.
    \end{equation} 
where we drop the potential $\pm$ sign of $\mc G^k(a)$.
\end{definition}

The task we are about to deal with is the following:

\begin{task}[Gate-dependent Pauli noise estimation]
    For a given multi-qubit Clifford gate $\mc G$, under the aforementioned assumptions, estimate the symmetrized Pauli eigenvalues $\bar\lambda_a^{\mc G}$ in a SPAM-robust manner.
\end{task}

We remark that learning the symmetrized Pauli eigenvalues is a standard task in the literature~\cite{erhard2019characterizing,hashim2020randomized}. By interleaving with carefully chosen single-qubit Clifford gates, one can learn more than those symmetrized eigenvalues~\cite{van2023probabilistic,chen2023learnability,berg2023techniques}. Such protocols can also be incorporated with \emeab{}, but we leave such demonstration for future work.
{Also note that, even learning those symmetrized Pauli eigenvalues is proved to cost exponentially many measurements for \sysb{} schemes~\cite[Theorem~2]{chen2024tight}.}

\medskip

\begin{proof}[Proof of correctness for \emeab{}]
    The protocol of \emeab{} has been outlined in details at Methods Sec B. To briefly recap, it consists of two subroutines, \eab{} and \auxb{}, which generate an estimator for $\lambda_a^\anc\bar{\lambda}_a$ and $\lambda^\anc_a$, respectively. Here we assume $\mc G$ is fixed and omit the corresponding superscript. The estimator of \emeab{} is then given by the ratio of these two estimators. Therefore, we just need to show both \eab{} and \auxb{} give consistent estimator. This again reduces to showing the fidelity estimator at depth $d$ is an unbiased estimator.

    \eab{}: Use $\mc C_j(\cdot)\coleq C_j(\cdot)C_j^\dagger$, $\mc P_j(\cdot)\coleq P_j(\cdot)P_j^\dagger$ to denote the corresponding unitary channel.
    Let us compute the probability of getting corrected measurement outcome $z$, averaged over random circuits at depth $d$,
    \begin{equation}\label{eq:eab_prob}
    \begin{aligned}
        \Pr[z] &=  
        \E \lbra{\widetilde\Phi_{z+\alpha+\beta}}\mc P_{\alpha,\beta}^{AS}\left(\mc C_{\mr{end}}\prod_{j=d}^1 \Lambda_\anc\mc C_j\right)^A \otimes \left(\mc P_{\mr{end}}\prod_{j=d}^1 \Lambda\mc G\mc P_j\right)^S\lket{\widetilde{\Phi}_+}\\
        &= \lbra{\Phi_{z}}\left(\E_{\alpha,\beta}\mc P_{\alpha,\beta}\mc E^\m\mc P_{\alpha,\beta}\right)^{AS}\left(\prod_{j=d}^1 \E_{\mc C_j'}\mc C_j^{'\dagger}\Lambda_\anc\mc C_j'\right)^A \otimes \left(\prod_{j=d}^1 \E_{\mc P_j'}\mc P_j'\Lambda\mc P_j'\mc G\right)^S\lket{\widetilde{\Phi}_+}\\
        &= \lbra{\Phi_z}\sum_{a_1,a_2\in\Pn}\xi_{a_1,a_2}^\m\lketbra{\sigma_{a_1,a_2}}{\sigma_{a_1,a_2}}^{AS}\left(\prod_{j=d}^1 \sum_{a_1\in\Pn}\lambda^\anc_{\pt(a_1)}\lketbra{\sigma_{a_1}}{\sigma_{a_1}} \right)^A \otimes \left(\prod_{j=d}^1 \sum_{a_2\in\Pn}\lambda_{a_2}\lketbra{\sigma_{a_2}}{\sigma_{a_2}}\mc G\right)^S\lket{\widetilde{\Phi}_+}\\
        &= \frac{1}{2^n}\sum_{a\in\Pn}(-1)^{\expval{a,z}}\xi_{a,a}^\m\lbra{\sigma_{a,a}}^{AS}\left( \sum_{a_1\in\Pn}(\lambda^\anc_{\pt(a_1)})^d\lketbra{\sigma_{a_1}}{\sigma_{a_1}} \right)^A \otimes \left(\sum_{a_2\in\Pn}(\lambda_{a_2}\lambda_{\mc G^\dagger(a_2)}\cdots\lambda_{\mc G^{\dagger d-1}(a_2)})\lketbra{\sigma_{a_2}}{\sigma_{a_2}}\right)^S\lket{\widetilde{\Phi}_+}\\
        &= \frac{1}{2^n}\sum_{a\in\Pn} (-1)^{\expval{a,z}}\xi_{a,a}^\m(\lambda_{\pt(a)}^\anc)^d\bar{\lambda}_a^d\lbraket{\sigma_{a,a}}{\widetilde{\Phi}_+}\\        
        &= \frac{1}{4^n}\sum_{a\in\Pn} (-1)^{\expval{a,z}}\xi_{a,a}^\m(\lambda_{\pt(a)}^\anc)^d\bar{\lambda}_a^d\xi_{a,a}^\s \\
        &\equiv \frac{1}{4^n}\sum_{a\in\Pn}(-1)^{\expval{a,z}} \xi_{a,a}(\lambda_{\pt(a)}^\anc)^d\bar{\lambda}_a^d.
    \end{aligned}
    \end{equation}
    
    \noindent   Recall that $\sigma_a = P_a/\sqrt{2^n}$ is the normalized Pauli operator, and we write $\sigma_{a_1,a_2}\coleq\sigma_{a_1}\otimes\sigma_{a_2}$.
    In the second line, we use the following change of variables
    \begin{equation}\label{eq:twirling_cov}
         C_j'\coleq  C_j C_{j-1}\cdots C_1,\quad  P_j'\coleq \mc G(P_j)\mc G^2(P_{j-1})\cdots\mc G^j(P_1),\quad\forall j=1,\cdots, d,
    \end{equation}
    which does not change the average thanks to the unitary invariance of the Haar measure. We also use the following fact of the noisy Bell measurement POVM
    \begin{equation}
    \begin{aligned}
        \lbra{\widetilde\Phi_{z+\alpha+\beta}} &= \frac{1}{2^n}\sum_{a\in\Pn}(-1)^{\expval{a,z+\alpha+\beta}}\lbra{\sigma_{a,a}}\mc E^\m \\
        &= \frac{1}{2^n}\sum_{a\in\Pn}(-1)^{\expval{a,z}}\lbra{\sigma_{a,a}}\mc P_{\alpha,\beta}\mc E^\m.
    \end{aligned}
    \end{equation}
    The third line uses the property of Pauli and local Clifford twirling (see e.g.~\cite{gambetta2012simulrb}). Note that Pauli twirling yields general Pauli channels, while local Clifford twirling yields Pauli channels of which the Pauli eigenvalues only depends on the pattern of the Pauli operator.

    From Eq.~\eqref{eq:eab_prob} and the Walsh-Hadamard transform, we see that $\hat f_a(d)[z]\coleq(-1)^{\expval{z,a}}$ is an unbiased estimator for $\xi_{a,a}(\lambda_{\pt(a)}^\anc)^d\bar{\lambda}_a^d$, for any $a\in\Pn$. Thus, by picking different $d$, averaging the estimators over different random circuits and measurement shots, then fitting to the single exponential decay model $\hat\xi_{a,a}^{\eab}({\hat\lambda^{\eab}_a})^d$, we obtain $\hat\lambda^{\eab}_a$ as an consistent estimator for $\lambda^\anc_{\pt(a)}\bar{\lambda}_a$.

    \medskip

    \auxb: Let us compute the probability of obtaining corrected outcome $z$ in \auxb{}, again averaged over random circuits at depth $d$.
    \begin{equation}\label{eq:AB_prob}
        \begin{aligned}
            \Pr[z] &= \E \lbra{\widetilde{z+\alpha_x}}\mc P_\alpha \mc C_{\mr{end}}\prod_{j=d}^1 \Lambda_\anc\mc C_j\lket{\widetilde{0}}\\
            &= \lbra{{z}}\left(\E_\alpha\mc P_\alpha\mc E^{'\m}\mc P_\alpha\right) \left(\prod_{j=d}^1 \E_{\mc C'_j}\mc C_j^{'\dagger}\Lambda_\anc\mc C'_j\right)\lket{\widetilde{0}}\\
            &= \lbra{{z}}\left(\sum_{a\in\Pn}\xi^{'\m}_{a}\lketbra{\sigma_a}{\sigma_a}\right) \left(\sum_{a\in\Pn}(\lambda_{\pt(a)}^\anc)^d\lketbra{\sigma_a}{\sigma_a}\right)\lket{\widetilde{0}}\\
            &= \frac1{2^n}\sum_{\mu\in\{0,1\}^n}(-1)^{z\cdot\mu}\xi_{\mu}^{'\m}\xi_{\mu}^{'\s}(\lambda_{\mu}^\anc)^d\\
            &\equiv \frac1{2^n}\sum_{\mu\in\{0,1\}^n}(-1)^{z\cdot\mu}\xi_{\mu}'(\lambda_{\mu}^\anc)^d
        \end{aligned}
    \end{equation}
    In the second line we do the same change of variables as Eq.~\eqref{eq:twirling_cov}, and use the following fact of noisy computational basis measurement
    \begin{equation}
        \begin{aligned}
            \lbra{\widetilde{z+\alpha_x}} = \lbra{{z+\alpha_x}}\mc E^{'\m} = \lbra{{z}}\mc X^{\alpha_x}\mc Z^{\alpha_z}\mc E^{'\m} = \bra{z}\mc P_\alpha\mc E^{'\m},
        \end{aligned}
    \end{equation}
    where we can add $\mc Z^{\alpha_z}$ as it acts trivially on the computational basis state. (Recall $\alpha_x,\alpha_z$ is the X and Z part of Pauli $P_\alpha$ as defined in Methods Sec A.)
    We also note that the SPAM noise channel here can be quite different from the \sysb{} experiment, as very different initial state and measurement settings are applied. 
    
    Thanks to Eq.~\eqref{eq:AB_prob} and the Walsh-Hadamard transform, we see that $\hat f_{\mu}(d)[z]\coleq(-1)^{z\cdot\mu}$ is an unbiased estimator for $\xi'_{\mu}(\lambda_{\mu}^\anc)^d$, for any $\mu\in\{0,1\}^n$. Thus, by picking different $d$, averaging the estimators over different random circuits and measurement shots, then fitting to the single exponential decay model $\hat\xi_{\mu}^{\auxb}({\hat\lambda^{\auxb}_\mu})^d$, we obtain $\hat\lambda^{\auxb}_\mu$ as an consistent estimator for $\lambda^\anc_{\mu}$. Choosing $\mu\coleq\pt(a)$ gives us the relevant $\lambda^\anc_{\pt(a)}$.
\end{proof}

\section{Validating formulas for \emeab{} variance}
In Methods Sec C, we obtain the following formulas for the variance overhead of \emeab{} over \sysb{}, assuming the use of ratio estimators,
\begin{equation}\label{eq:up_to_here_works}
    \frac{\mr{Var}[\hat\lambda_\emeab]}{\mr{Var}[\hat\lambda_{\sysb}]}\approx 
    \frac{d_\sysb^2\alpha_\sysb^2V_\eab}{d_\eab^2\alpha_\eab^2V_\sysb} + \frac{d_\sysb^2\alpha_\sysb^2V_\auxb}{d_\auxb^2\alpha_\auxb^2V_\sysb}
\end{equation}
where $V_\eab{}$ is defined as (similar for $V_\auxb{}$, $V_\sysb{}$)
\begin{equation}
    V_\eab\coleq \mr{Var}[\hat f_0] + \mr{Var}[\hat f_d]/\lambda_{\eab{}}^{2d},
\end{equation}
where $\hat f_m$ is defined as the fidelity estimator at depth $m\in\{0,d\}$ for the corresponding protocols.

\medskip

By further assuming that (1) \eab{}, \auxb{}, \sysb{} all use a ratio estimator at their optimal circuit depth~\cite{harper2019statistical}; (2) The fidelity estimator at every depth satisfies a binomial variance; (3) The noise are sufficiently weak to allow a first-order approximation, we obtain the following formula
\begin{equation}\label{eq:final_overhead}
\begin{aligned}
    \frac{\mr{Var}[\hat\lambda_\emeab]}{\mr{Var}[\hat\lambda_{\sysb}]}&\approx
    \left(1+\frac{r_\auxb}{r_\sysb{}}\right)^2\frac{\alpha_\sysb^2}{\alpha^2_\eab{}} + 1
\end{aligned}
\end{equation}
which roughly scales as $\alpha_\sysb^2/\alpha^2_\eab$, i.e., square of the SPAM fidelity ratio. 

In Fig.~4 in the main text, we calculated the overhead from experimental data by comparing the variance of the \emeab{} and \sysb{} extracted fidelities as reported by scipy's \textsf{curve\_fit} function~\cite{2020SciPy-NMeth}. Specifically, we examined the reported variance of the extracted \sysb{}, \auxb{}, and \eab{} fidelities by fitting  $f_d = \alpha \lambda^d$ (dropping the model subscript for brevity), where the variance of individual data points are supplied to the \textsf{curve\_fit} function. We then calculate the variance of \emeab{} by linearly expanding the variance of $\hat\lambda_{\eab{}}/\hat\lambda_{\auxb{}}$.  We showed that the theoretically calculated overhead at optimal depth~\eqref{eq:final_overhead} captures the correct trend of the experimentally extracted overhead, but observed a difference between the exact values. We attributed those differences to the violation of assumptions (1) and (2) in deriving Eq.~\eqref{eq:final_overhead}. 

\begin{figure}
    \centering
    \includegraphics[width=0.5\textwidth]{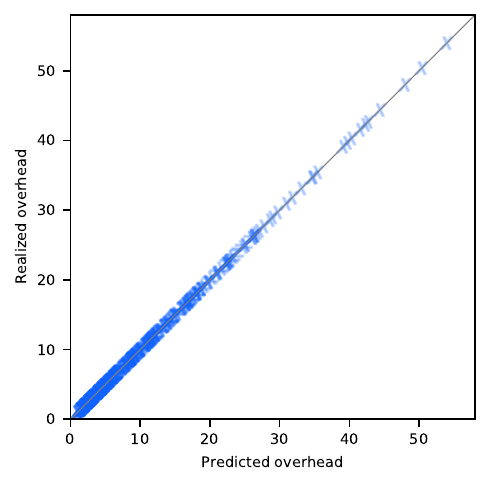}
    \caption{\textbf{Comparison of overhead estimates.} 
    Comparing the overhead obtained from performing the fit using only $d=0,32$ from the experiment in Fig.~4 in the main text with \textsf{curve\_fit} labeled as ``Realized overhead" versus predictions of Eq.~\eqref{eq:up_to_here_works} labeled as ``Predicted overhead". The data falls on the 45 degree line, showing perfect agreement between the two estimates and validating Eq.~\eqref{eq:up_to_here_works}.}
    \label{fig:comp-var}
\end{figure}

Here, we show that the overhead from performing the fit using only $d=0,32$ matches the estimated overhead using the ratio estimator in Eq.~\eqref{eq:up_to_here_works}. In Fig.~\ref{fig:comp-var}, we plot the overhead obtained from \textsf{curve\_fit} labeled as ``Realized overhead" versus predictions of Eq.~\eqref{eq:up_to_here_works} labeled as ``Predicted overhead". We observe that, indeed, the data falls on the 45 degree line, showing perfect agreement between the two estimates and validating Eq.~\eqref{eq:up_to_here_works}.

\section{Context-aware dynamical decoupling}
To suppress correlated errors between system and auxiliary qubits, as well as to reduce errors on the latter, we employ the context-aware dynamical decoupling (CA-DD) scheme introduced in Ref.~\cite{seif2024suppressing}. This scheme considers the context of the circuit, including the configuration of gates and idling periods, to insert dynamical decoupling (DD) pulses.

\begin{figure}
    \centering
    \includegraphics[width=0.61\textwidth]{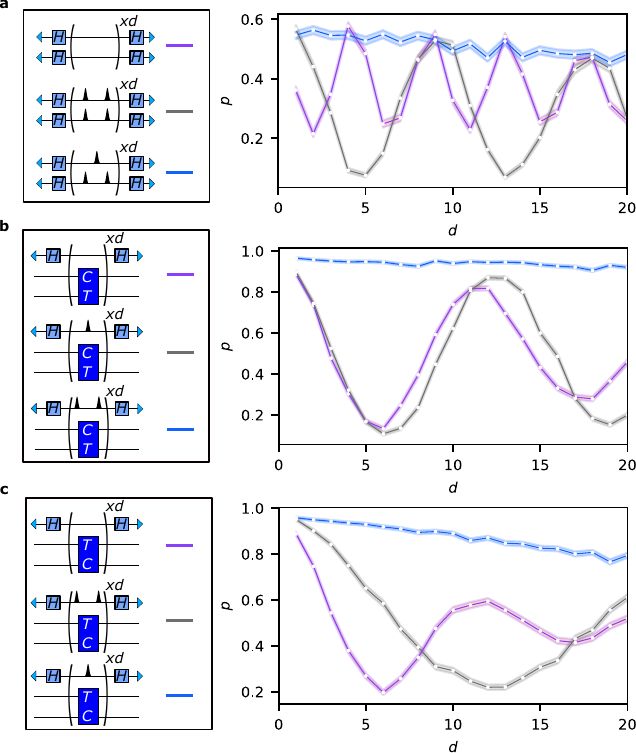}
    \caption{\textbf{Context-aware dynamical decoupling.} Ramsey experiments in various contexts with different dynamical decoupling (DD) sequences. Pulse sequences are repeated $d$ times (left panels) and the probability of return to the initial state $p$ is measured, which should ideally remain at 1. Due to coherent errors between qubits, $p$ oscillates. The purple curves show noisy results without any DD, the grey curves indicate context-unaware DD with aligned pulses, and the blue curves represent context-aware DD (CA-DD) with staggered pulses. In all cases, CA-DD performs the best and suppresses correlated coherent noise.
    \textbf{a,} Two adjacent idle qubits experiencing correlated $ZZ$ error. Each interval is an idle period of 500 ns. \textbf{b,} The control spectator qubit, and \textbf{c,} The target spectator qubit of the echoed-cross resonance gate. The experiments are performed on \emph{ibm\_nazca} on qubits (56, 57) with 5000 shots in panel (a), on qubit 53 with the gate on qubits (60, 59) with 2000 shots in panel (b), and on qubit 53 with the gate on qubits (61, 60) with 2000 shots in panel (c).}
    \label{fig:ca_dd}
\end{figure}

A major source of correlated noise on the experimental systems used in this work is the static $ZZ$ error that arises from coupling to higher levels of the transmon qubits~\cite{Tripathi_Operation_2019, Magesan_Effective_2020, Malekakhlagh_First_2020}. The Hamiltonian of the error on a pair of qubits is given by
\begin{equation}\label{eq:H11}
    H_{11} = \frac{\nu}{2} (-I\otimes Z - Z\otimes I + Z\otimes Z),
\end{equation}
where $\nu$ is the strength of the coupling. This type of error between arbitrary pairs of qubits can be effectively eliminated by utilizing staggered DD sequences~\cite{leung2002simulation,rotteler2006equivalence,zhou2022quantum,Shirizly2024dissipative,seif2024suppressing}. The simplest example of staggered pulses used to suppress this error is shown in Fig.~\ref{fig:ca_dd}a, where we apply $X$ pulses on the idle qubits, such that the idle times between pulses are staggered. Specifically, we use $S_1 = I_{\tau/4}$-$X$-$I_{\tau/2}$-$X$-$I_{\tau/4}$ and $S_2 = I_{\tau/2}$-$X$-$I_{\tau/2}$-$X$, where $I_{t}$ is the idle evolution for a period of $t$. Note that the pulses in $S_2$ are shifted by $\tau/4$ compared to $S_1$. For a total idle period of $\tau$, these sequences suppress the $ZZ$ error since 
\begin{align}\label{eq:staggDD}
    U_\tau &=e^{-i\tau/4 H_{11}}.IX.e^{-i\tau/4 H_{11}}.XI.e^{-i\tau/4 H_{11}}.IX.e^{-i\tau/4 H_{11}}.XI \\
    &= e^{-i\tau/4 H_{11}}.e^{+i\tau/4 H_{11}}.e^{-i\tau/4 H_{11}}.e^{+i\tau/4 H_{11}}\\&=I
\end{align}

The presence of gates can modify the error process, making it crucial to consider the pulses implementing the gate when applying DD. In our experiments, the native two-qubit gate, the echoed cross-resonance (ECR) gate, includes intrinsic DD pulses. These pulses must be taken into account when applying DD to idle qubits adjacent to active qubits where the gate is being applied. Specifically, the ECR gate includes an echo pulse ($X$) on its control qubit and two rotary echo pulses on its target qubit. Therefore, the simplest DD pulses on idle qubits that correctly stagger with the pulses on active qubits are represented by $S_1$ ( Fig.\ref{fig:ca_dd}b) and $S_2$ ( Fig.\ref{fig:ca_dd}c) sequences for the control and target spectators, respectively.

In this work,  we focus on nearest-neighbor crosstalk errors and employ CA-DD to effectively suppress them using the fewest possible pulses. For the layouts we considered, it turns out that only two distinct pulses ($S_1$ and $S_2$) are required to suppress the crosstalk errors. 

To illustrate the importance of DD, we examine the three cases mentioned above in experiments. Specifically, we perform Ramsey experiments under various scenarios and probe the coherence of qubits with different DD pulse configurations. Across all experiments, properly staggered CA-DD sequences effectively suppress both single-qubit errors and two-qubit correlated errors. In contrast, improperly selected DD pulses only address single-qubit errors and fail to suppress correlated errors.

First, we examine two idling qubits. We initialize the qubits in the $\ket{++}$ state and allow them to evolve under three distinct scenarios: (1) without DD, (2) with simple independent DD ($S_1$ on both qubits), and (3) with CA-DD and correct staggering  ($S_1$ and $S_2$ on the qubits). We then measure $p$ the probability of finding qubits in the $\ket{++}$ state, which should ideally remain at 1. As shown in Fig.~\ref{fig:ca_dd}a, staggered pulses yield the optimal performance, while independent DD pulses fail to fully suppress the errors.

In the second case, we focus on the control spectator and conduct single-qubit Ramsey experiments on the idle qubit while the gate is applied to its neighboring qubits. We initialize the spectator qubit in the $\ket{+}$ state and observe its evolution under three different scenarios: (1) without DD, with incorrectly aligned DD pulse ($S_1$), and (3) CA-DD with correct staggering in the context  ($S_2$). We then measure the probability $p$ of finding the qubits in the $\ket{+}$ state, which should ideally remain at 1. As illustrated in Fig.~\ref{fig:ca_dd}b, staggered pulses yield the best performance, while the application of an incorrect DD pulse fails to suppress the correlated errors.

Finally, we focus the target spectator and conduct a similar single qubit Ramsey experiment. In this case, the correctly staggered sequence is $S_1$, and the incorrectly aligned sequence is $S_2$. 
As shown in Fig.~\ref{fig:ca_dd}c, staggered pulses yield the best performance, whereas the application of an incorrect DD pulse fails to suppress the correlated errors.

We remark that in our experiments, when implementing the $S_2$ sequence, we exclude the final $X$ pulse. This exclusion is possible as the pulse can be absorbed within the twirling gates in the learning experiment.

\section{Experimental details and additional results}
In this section we present the details of experiments performed in this work. We also provide additional results concerning learning noise on gates in the experiments. 

\subsection{Overview of devices}

In the characterization of a quantum processor, understanding the error properties and coherence times is important for understanding the limits and  reliability of the device. The key parameters typically reported include the gate fidelity, readout error, and coherence times (T1 and T2). 

Gate fidelity measures the accuracy with which our quantum gates operate. Here, we report the average gate fidelity measured through standard randomized benchmarking. For a given gate, the fidelity can then be thought of as an average over all possible input states of the fidelity between the state produced by the actual operation and the state produced by the ideal operation. For a noise channel $\Lambda$ the average fidelity then is for some state $\rho_{\psi}=\ketbra{\psi}{\psi}$ defined over the uniform Haar average,
\begin{align}
F_{\mathrm{avg}}\left(\Lambda\right) & =\mathbb{E}_{\psi}\left[F\left(\rho_{\psi},\Lambda\left(\rho_{\psi}\right)\right)\right]\label{eq:FavgGateFidelity}\\
 & =\int\mathrm{d}\psi\,\braOket{\psi}{\Lambda\left(\kb{\psi}{\psi}\right)}{\psi}\nonumber \\
 & =\int\mathrm{d}\psi\,\bb\rho_{\psi}\vert\Lambda\vert\rho_{\psi}\kk\;.\nonumber 
\end{align}

Readout errors indicates the probability of measurement inaccuracies. 
The average assignment readout error is \(1 - \mathcal{F}_a\), where \(\mathcal{F}_a\) denotes the readout assignment fidelity of a qubit, defined as:
\[
\mathcal{F}_a = 1 - \frac{1}{2} \left( P(1|0) + P(0|1) \right)\;.
\]
Here, \(P(A|B)\) represents the empirical probability of measuring the qubit in state \(A \in \{0,1\}\) given that it was nominally prepared in state \(B \in \{0,1\}\). 

Finally, the coherence times, T1 (energy relaxation time) and T2 (phase coherence time) are the key measures of how long a qubit can maintain its quantum state before losing information.

\begin{figure}[t]
    \centering 
\includegraphics[width=1.0\textwidth]{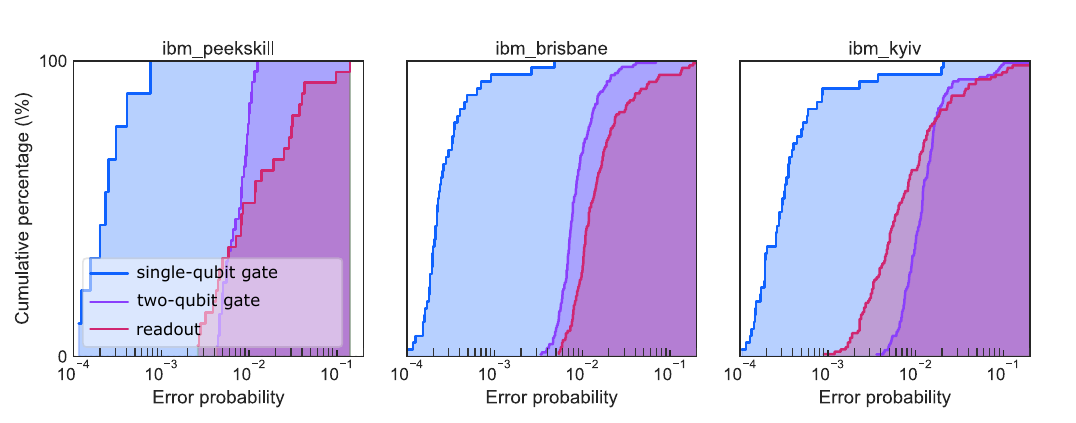}
\caption{
Cumulative distribution functions (CDFs) for error probabilities observed in three IBM quantum devices: \textit{ibm\_peekskill}, \textit{ibm\_brisbane}, and \textit{ibm\_kyiv}. Shaded regions under the CDF curves illustrate the distribution of error probabilities across the devices. The graphs present error probabilities for single-qubit gates, two-qubit gates, and readout errors. Single-qubit gate errors refer to those of the square root of an X gate ($\sqrt{X}$), while two-qubit gate errors are for the native ECR gate. The x-axis depicts error probability on a logarithmic scale, and the y-axis represents the cumulative percentage of qubits or gates.
}
 \label{fig:si_device_error_rates}
\end{figure}

\begin{figure}[t]
    \centering 
\includegraphics[width=1.0\textwidth]{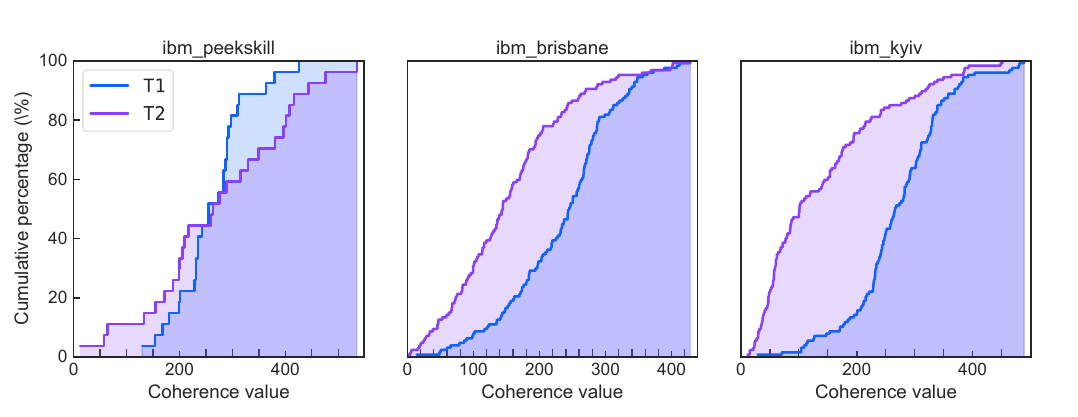}
\caption{
Cumulative distribution functions (CDFs) of the energy relaxation $T_1$ and Hanh-echo coherence $T_2$ times for three IBM quantum devices: \textit{ibm\_peekskill}, \textit{ibm\_kyiv}, and \textit{ibm\_kyiv}. The charts illustrate the cumulative percentage of qubits exhibiting specific T1 and T2 coherence times. The x-axis indicates the coherence time (in $\mu$s), and the y-axis displays the cumulative percentage of qubits.
}
\label{fig:si_device_coherence}
\end{figure}

Figures \ref{fig:si_device_error_rates} and \ref{fig:si_device_coherence} depict the empirical cumulative distribution functions (CDFs) for these error probabilities and coherence times, respectively, for three IBM quantum devices: \textit{ibm\_peekskill}, \textit{ibm\_brisbane}, and \textit{ibm\_kyiv}. 

Table \ref{tab:error_rates} summarizes the error probabilities for single-qubit gates, two-qubit gates, and readout errors for the three devices. Table \ref{tab:coherence_times} provides the coherence times (T1 and T2) for the three devices.

\begin{table}[h]
    \centering
    \begin{tabular}{|c|c|c|c|c|c|c|}
        \hline
        \textbf{Device} & \multicolumn{2}{c|}{\textbf{Single-Qubit Gate Error}} & \multicolumn{2}{c|}{\textbf{Two-Qubit Gate Error}} & \multicolumn{2}{c|}{\textbf{Readout Error}} \\
        \hline
        & \textbf{Mean} & \textbf{Median} & \textbf{Mean} & \textbf{Median} & \textbf{Mean} & \textbf{Median} \\
        \hline
        \textit{ibm\_peekskill} & $2.8 \times 10^{-4}$ & $2.3 \times 10^{-4}$ & $7.6 \times 10^{-3}$ & $7.9 \times 10^{-3}$ & $2.2 \times 10^{-2}$ & $8.5 \times 10^{-3}$ \\
        \hline
        \textit{ibm\_brisbane} & $4.4 \times 10^{-4}$ & $2.2 \times 10^{-4}$ & $1.0 \times 10^{-2}$ & $7.8 \times 10^{-3}$ & $2.4 \times 10^{-2}$ & $1.2 \times 10^{-2}$ \\
        \hline
        \textit{ibm\_kyiv} & $1.4 \times 10^{-3}$ & $3.0 \times 10^{-4}$ & $2.3 \times 10^{-2}$ & $1.2 \times 10^{-2}$ & $1.7 \times 10^{-2}$ & $6.4 \times 10^{-3}$ \\
        \hline
    \end{tabular}
    \caption{Error probabilities for single-qubit gates, two-qubit gates, and readout errors for \textit{ibm\_peekskill}, \textit{ibm\_brisbane}, and \textit{ibm\_kyiv}.}
    \label{tab:error_rates}
\end{table}

\begin{table}[h]
    \centering
    \begin{tabular}{|c|c|c|c|c|}
        \hline
        \textbf{Device} & \multicolumn{2}{c|}{\textbf{T1 Time} ($\mu$s)} & \multicolumn{2}{c|}{\textbf{T2 Time} ($\mu$s)} \\
        \hline
        & \textbf{Mean} & \textbf{Median} & \textbf{Mean} & \textbf{Median} \\
        \hline
        \textit{ibm\_peekskill} & 260 & 250 & 270 & 260 \\
        \hline
        \textit{ibm\_brisbane} & 230 & 240 & 150 & 140 \\
        \hline
        \textit{ibm\_kyiv} & 270 & 270 & 140 & 100 \\
        \hline
    \end{tabular}
    \caption{Coherence times (T1 and T2) for \textit{ibm\_peekskill}, \textit{ibm\_brisbane}, and \textit{ibm\_kyiv}.}
    \label{tab:coherence_times}
\end{table}

In all experiments, we utilize a fixed decomposition for the twirling gates to minimize the variations in single qubit gate errors. Specifically, we decompose the Pauli and Clifford gates required for twirling as 
\begin{equation}
    U = R_z(\alpha+\pi)\sqrt{X}R_z(\beta+\pi)\sqrt{X}R_z(\gamma), 
\end{equation}
where $\alpha,\beta,\gamma$ are the Euler angles, as shown in Figs.~\ref{fig:inject_circs},~\ref{fig:pair_circs},~and~\ref{fig:layer_circs}.  

We also employ CA-DD in all experiments expect the hypothesis testing example in Fig.~2 of the main text. This ensures that the correlations between system and auxiliary qubits are suppressed. The agreement between \emeab{} and \sysb{} results further validate the success of CA-DD in suppressing such correlations. 

In the following we present further details and example circuits for the experiments featured in Figs.~2,~3,~and~4 in the main text and include additional results supplementing those in Figs.~3,~and~4 in the main text. 

\subsection{Hypothesis testing}
\begin{figure}[t]
    \centering
    \includegraphics[width=\textwidth]{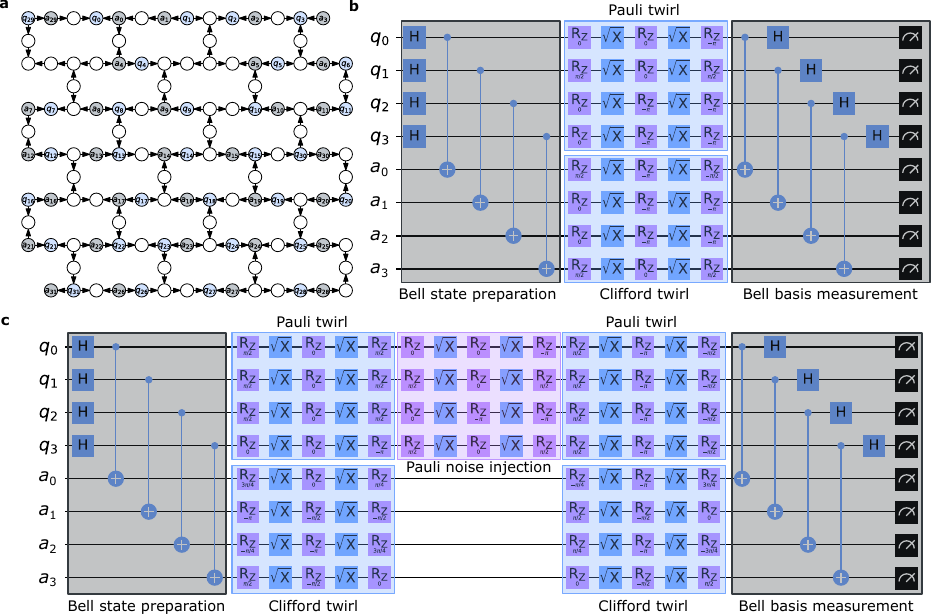}
    \caption{\textbf{Hypothesis testing layout and circuits.} \textbf{a,} The layout used for the experiments in Fig.~2 in the main text on \emph{ibm\_brisbane}. System qubits are highlighted in blue, while auxiliary qubits are in grey. Different system sizes in Fig.~2 use qubits in this layout in increasing order.  \textbf{b,} A depth-0 \eab{} circuit with 4 system qubits and 4 auxiliary qubits. The system qubits are Pauli twirled and auxiliary qubits are local Clifford twirled. The grey boxes indicate state preparation and measurement in the Bell basis as required by the \eab{} protocol. \textbf{c,} A depth-1 \eab{} circuit including the twirling layer and a noise injection layer. The  noise is injected by randomly inserting a Pauli operator between the twirling layers. }
    \label{fig:inject_circs}
\end{figure}

The first experiment concerning hypothesis testing was performed on \emph{ibm\_brisbane} using up to 64 qubits. In Fig.~\ref{fig:inject_circs}a we show the layout used in Fig.~2 in the main text. We consider system sizes $n=4$ to 32 qubits in increments of 4. The layout for different problem sizes correspond to qubits chosen in increasing order starting from $q_0$ and $a_0$ for system and auxiliary qubits, respectively. In this example we used $d=0$ (Fig.~\ref{fig:inject_circs}b) and $d=1$ (Fig.~\ref{fig:inject_circs}c) to obtain a SPAM robust estimate of the fidelity of the injected noise. At each depth and size we first sample 1000 random twirls each with 10000 shots. Each random twirl at $d=1$ is accompanied by an additional random inserted Pauli as shown in Fig.~\ref{fig:inject_circs}c.

We then use the data and the information about injected Paulis to emulate a desired Pauli channel. Specifically, in case of the depolarization channel $
    \Lambda_D(\hat\rho) = \frac{1}{2^n} [\hat{I} \Tr(\hat\rho)]$, 
we randomly choose 500 out of the 1000 inserted Pauli operations. Note that for a given $P_k$, the 500 randomly chosen Pauli operations either commute or anti-commute with $P_k$ with equal probability. Therefore, on average we have $\lambda_k=0$. 

For the spiked Pauli channels     $\Lambda_\pm(\hat\rho) = \frac{1}{2^n} [\hat{I} \Tr(\hat\rho) \pm \frac{1}{3} \hatpk \Tr(\hat\rho \hatpk)]$, we first choose $P_k$ and $\lambda_k = \pm 1/3$ at random. Note that for a given $\lambda_k$, the probability of having a  Pauli error that commutes with $P_k$ is given by $p_k=(1+\lambda_k)/2$. Therefore, we analyze the list of circuits with inserted random Paulis and label them to indicate whether the inserted Pauli operation commutes or anticommutes with $P_k$. We then find the number of commuting Paulis $N_k$ according to a Binomial distribution with 500 trials and $p_k$, and choose $N_k$ Paulis at random from the set of inserted Pauli operators that commute with $P_k$. Finally, we sample the rest of $500-N_k$ Paulis from anticommuting ones. This procedure ensures that the emulated channels have the fidelity $\lambda_k=\pm1/3$ on average as desired. 

We repeat the above procedure 50 times for both depolarization and spiked Pauli channels to obtain the data presented in Fig.~2 in the main text. 

\subsection{Learning noise in a single two-qubit gate}
\begin{figure}[h]
    \centering
    \includegraphics[width=\textwidth]{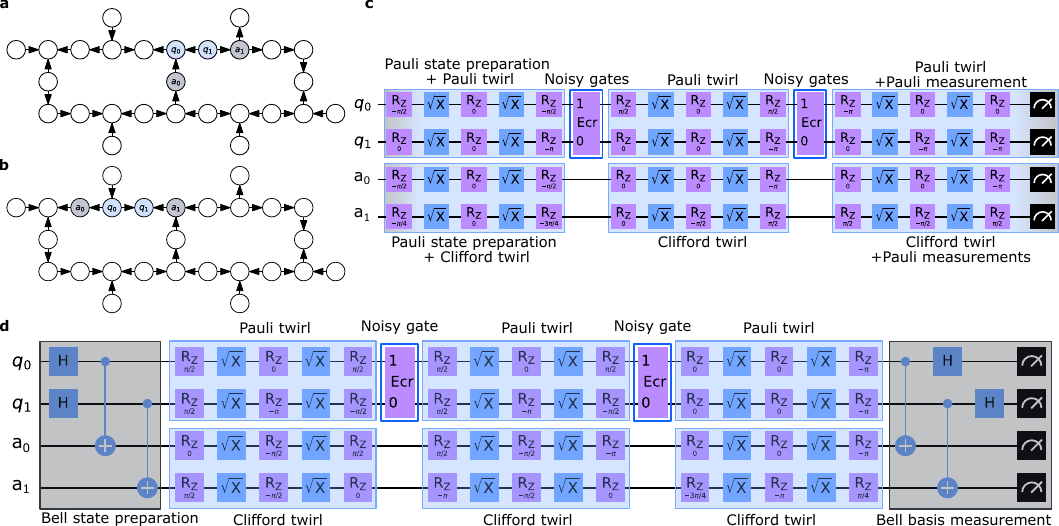}
    \caption{\textbf{Layouts and circuits for learning noise on a two-qubit gate.} \textbf{a,} The layout used for the experiments in Fig.~3 in the main text on \emph{ibm\_peeskill}. System qubits are highlighted in blue, while auxiliary qubits are in grey.   \textbf{b,} A different layout used for experiments shown in Fig.~\ref{fig:pair_fids}. \textbf{c, } A depth-2 \sysb{} and \auxb{} circuit with 2 system qubits and 2 auxiliary qubits. The system qubits are Pauli twirled and auxiliary qubits are local Clifford twirled.  The noisy echoed-cross resonance (ECR) gate is applied to the system. The state-preparation and measurement operations only require single-qubit gates, which are combined with the twirling layers.   \textbf{d,} A depth-2 \eab{} circuit.
    The grey boxes indicate state preparation and measurement in the Bell basis as required by the \eab{} protocol. The twirling layers and noisy gate layers are similar to the previous panel.}
    \label{fig:pair_circs}
\end{figure}
The experiment concerning noise in a single two-qubit gate was performed using 4 qubits on \emph{ibm\_peekskill}. In Fig.~\ref{fig:pair_circs}a we show the layout used in Fig.~3 in the main text. In Fig.~\ref{fig:pair_circs}c, we show the circuits used for \sysb{} (upper half) and \auxb{} (lower half). In Fig.~\ref{fig:pair_circs}d, we show the circuits used for \eab{}. Note that in these figures, we choose $d=2$ to simplify the presentation. However, the data presented in Fig.~3 were obtained using $d=0,16,32$ with 100 twirls, each with 1000 shots.

\begin{figure}[h]
    \centering
    \includegraphics[width=0.5\textwidth]{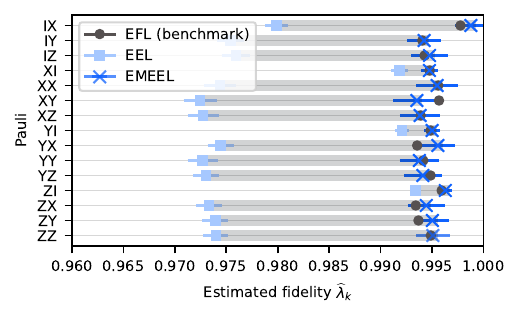}
    \caption{\textbf{Learning noise on a two-qubit gate.} Estimated Pauli fidelities $\hat\lambda_k$   of $\Lambda_s$ for the 15 non-trivial Pauli ($P_k$) operators on the two qubit depicted in Fig.~\ref{fig:pair_circs}b on \emph{ibm\_peekskill}. Similar to Fig.~3 in the main text, we measure the fidelities using two independent methods with and without quantum memory. We take \sysb{} as the ground truth. Without error mitigation \eab{} results do not match \sysb{}'s, with their disagreement highlighted in grey.  However, after applying error mitigation we uncover correct fidelities with \emeab{}. In this layout one the auxiliary qubits have a much better fidelity than the other one as demonstrated by the difference between the single-qubit \eab{} fidelities. Error bars are statistical correspond to 1.96 standard deviation.}
    \label{fig:pair_fids}
\end{figure}

We also consider a different layout on  \emph{ibm\_peekskill} for this example (see Fig.~\ref{fig:pair_circs}b). As shown in Fig.~\ref{fig:pair_fids}, we again observe that while \eab{} results deviate form the benchmark \sysb{} results, using error mitagtion \emeab{} recovers the correct fidelities. Moreover, we observe that the noise on the two auxiliary qubits is very different in this experiment, resulting in a small deviations of \eab{} fidelity estimates from \sysb{}'s in one of the qubits, and a larger deviation in the other one. 

\subsection{Learning noise in a layer of two-qubit gate}
\begin{figure}[t]
    \centering
    \includegraphics[width=0.95\textwidth]{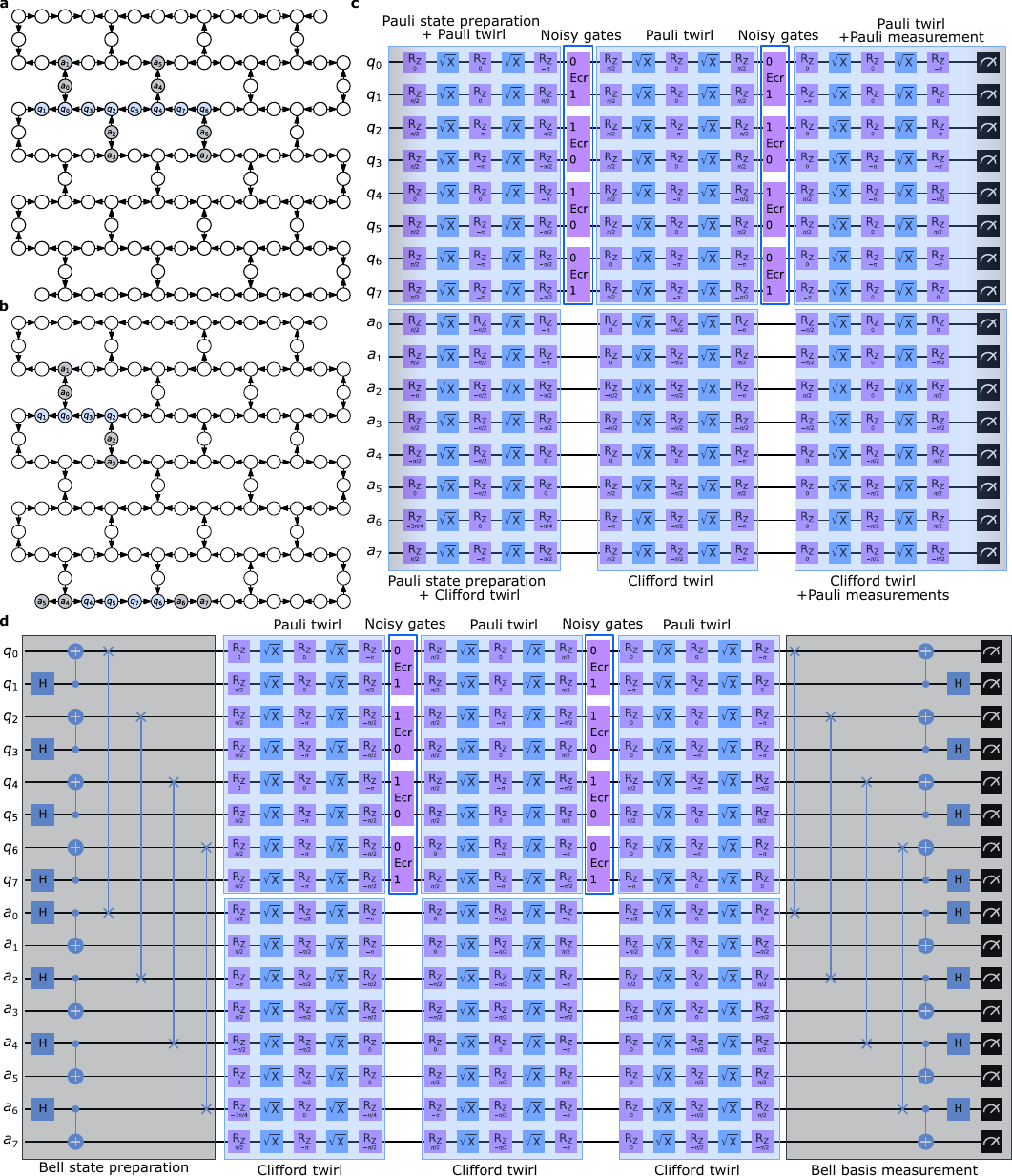}
    \caption{\textbf{Layouts and circuits for learning noise on layer of 4 two-qubit gate.} \textbf{a,} The layout used for the experiments in Fig.~4 in the main text on \emph{ibm\_kyiv}. System qubits are highlighted in blue, while auxiliary qubits are in grey.   \textbf{b,} A different layout used for experiments shown in Fig.~\ref{fig:far_fids}. \textbf{c, } A depth-2 \sysb{} and \auxb{} circuit with 8 system qubits and 8 auxiliary qubits. The system qubits are Pauli twirled and auxiliary qubits are local Clifford twirled.  The noisy echoed-cross resonance (ECR) gates are applied to the system qubits. The state-preparation and measurement operations only require single-qubit gates, which are combined with the twirling layers. \textbf{c,} A depth-2 \eab{} circuit.
    The grey boxes indicate state preparation and measurement in the Bell basis as required by the \eab{} protocol. Compared to Fig.~\ref{fig:pair_circs} there are additional SWAP gates needed to entangle nonadjacent system-auxiliary qubit pairs. Twirling layers and noisy gate layers are similar to the previous panel.}
    \label{fig:layer_circs}
\end{figure}
The experiment concerning noise in a layer of 4 two-qubit gate was performed using 16 qubits on \emph{ibm\_kyiv}. In Fig.~\ref{fig:layer_circs}a we show the layout used in Fig.~4 in the main text. In Fig.~\ref{fig:layer_circs}c, we show the circuits used for \sysb{} (upper half) and \auxb{} (lower half). In Fig.~\ref{fig:layer_circs}d, we show the circuits used for \eab{}. Note that in these figures, we choose $d=2$ to simplify the presentation. However, the data presented in Fig.~3 in the main text were obtained using $d=0,32$ with 500 twirls, each with 500 shots.

We also consider a different layout on  \emph{ibm\_kyiv} for this example with two disjoint group of qubits (Fig.~\ref{fig:layer_circs}b). In this layout $\{q_0,q_1,q_2,q_3\}$ are in one group, and $\{q_4,q_5,q_6,q_7\}$ are in a different group distant from the first one. Therefore, we have 4 pair of qubits such that the first two pairs, i.e., $(q_0,q_1)$ and $(q_2,q_3)$, are adjacent, and the other two pairs, i.e., $(q_4,q_5)$and $(q_6,q_7)$, are adjacent to each other and distant from the first two. 
\begin{figure}[t]
    \centering
    \includegraphics[width=\textwidth]{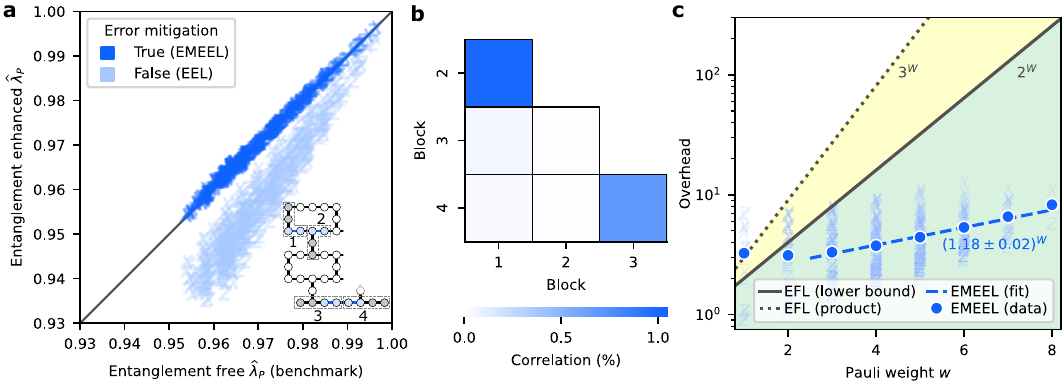}
    \caption{\textbf{Learning noise in a layer of parallel gates.} Similar to the experiment depicted in Fig.4 of the main text, but conducted on a distinct layout featuring two distant groups of qubits, as illustrated in Fig.\ref{fig:layer_circs}b.  \textbf{a, } The estimated Pauli fidelities $\hat \lambda_P$ using entanglement enhancement with quantum memory (\eab{}) and  entanglement-free learning without quantum memory   (\sysb{}) should ideally align with the 45-degree line, as they correspond to the same physical quantity derived from two independent methods. However, due to bias from auxiliary noise, they do not. Error-mitigated entanglement-enhanced learning (\emeab{}) restores the correct values by using information from \auxb{} to mitigate errors in \eab{} estimates. In principle,  \emeab{} enables access to all the Pauli fidelities, however, due to the prohibitive cost of \sysb{} we only benchmark them in the $X$ and $Z$ basis (markers). The inset shows the layout.
    \textbf{b, } The average correlation between two-qubit $XX$ and $ZZ$ fidelities of each pair of blocks. Adjacent pairs exhibit strong correlation (1,2) and (3,4), whereas distant pairs show negligible correlations. The  The block index $i+1$ corresponds to qubit pairs  as shown in the inset of panel a.
    \textbf{c, } The overhead of \emeab{} compared with a theoretical lower bound for \sysb{} as a function of the Pauli weight $w$, i.e., the number of non-trivial qubits that the operator acts on. The lower overhead of \emeab{}, $(1.18\pm0.02)^w$ from fit, compared to the $2^w$ theoretical lower bound for \sysb{} and the $3^w$ overhead for single-qubit measurements demonstrate entanglement enhancement in sample complexity.}
    \label{fig:far_fids}
\end{figure}

As shown in Fig.~\ref{fig:far_fids}a, we again observe that while \eab{} results deviate form the benchmark \sysb{} results, using error mitigation \emeab{} recovers the correct fidelities. Moreover, we observe that the correlations are strong between adjacent pairs and almost non-existent between distant pairs (Fig.~\ref{fig:far_fids}b). This experiment was performed with $d=0,32$ and 
444 twirls with 500 shots each. Finally, we observe that in this layout, the overhead factor of \emeab{} is $(1.18\pm0.02)^n$, which is much smaller than the $2^n$ theoretical lower bound for \sysb{}.

\bibliographystyle{naturemag}
\bibliography{paulilearning}